\documentclass[journal]{IEEEtran}
\usepackage{makeidx}  
\usepackage{cite}
\usepackage{graphicx}
\usepackage{mdwtab}
\usepackage{algorithm}
\usepackage{algorithmic}
\usepackage{amsmath}
\usepackage{tabularx}
\usepackage{dirtree}
\usepackage{tikz}
\usepackage{color}
\usepackage{caption}
\usepackage{subcaption}

\usepackage[a4paper,left=15mm,right=15mm, top=1cm, bottom=2cm]{geometry}

\begin{document}

	\title{A Comprehensive Survey on Moving Networks}
	
	\author{\IEEEauthorblockN{Shan Jaffry$^1$, Rasheed Hussain$^2$, Xiang Gui$^3$, Syed Faraz Hasan$^3$}\\
		\IEEEauthorblockA{
			$^1$DGUT-CNAM Institute, Dongguan University of Technology, Dongguan, China.\\
			E: S.Jaffry@dgut.edu.cn\\
			$^2$Institute of Information Security and Cyber-Physical Systems, Innopolis University, Russia\\
			E: r.hussain@innopolis.ru\\
			$^3$Department of Mechanical and Electrical Engineering, Massey University, New Zealand.\\
			E \{X.Gui, F.Hasan\}@massey.ac.nz
			}
	}
	
	\maketitle
	
\begin{abstract}
The unprecedented increase in the demand for mobile data, fuelled by new emerging applications and use-cases such as high-definition video streaming and heightened online activities has caused massive strain on the existing cellular networks. As a solution, the fifth generation (5G) of cellular technology has been introduced to improve network performance through various innovative features such as millimeter-wave spectrum and Heterogeneous Networks (HetNet). In essence, HetNets include several small cells underlaid within macro-cell to serve densely populated regions like stadiums, shopping malls, and so on. Recently, a mobile layer of HetNet has been under consideration by the researchers and is often referred to as moving networks. Moving networks comprise of mobile cells that are primarily introduced to improve Quality of Service (QoS) for commuting users inside public transport because the QoS is deteriorated due to vehicular penetration losses and high Doppler shift. Furthermore, the users inside fast moving public transport also exert excessive load on the core network due to large group handovers. To this end, mobile cells will play a crucial role in reducing overall handover count and will help in alleviating these problems by decoupling in-vehicle users from the core network. This decoupling is achieved by introducing separate in-vehicle access link, and out-of-vehicle backhaul links with the core network. Additionally sidehaul links will connect mobile cells with their neighbors.

To date, remarkable research results have been achieved by the research community in addressing challenges linked to moving networks. However, to the best of our knowledge, a discussion on moving networks and mobile cells in a holistic way is missing in the current literature. To fill the gap, in this paper, we comprehensively survey moving networks and mobile cells. We cover the technological aspects of moving cells and their applications in the futuristic applications. We also discuss the use-cases and value additions that moving networks may bring to future cellular architecture and identify the challenges associated with them. Based on the identified challenges we discuss the future research directions. 
	\end{abstract}
	

\section{Introduction}

Cellular networks, being most popular wireless communication mode, are facing an unprecedented rise in data demands driven by applications, such as, 4K/8K video streaming, wireless augmented reality (AR) based gaming, video sharing or callings through instant messaging applications, etc. \cite{akamai2017report,inoue2017field}. According to Ericsson, the mobile data demand will exceed 109 Exa Bytes (Exa = $10^{18}$) per month by 2023 \cite{cerwall2018ericsson} which was only 20 Exa Bytes in 2018. Current cellular networks needs serious paradigm shift in order to meet these increasing traffic requirements.
 
The fifth generation (5G) and beyond networks will change the current network paradigm by introducing several new features, such as, millimeter wave (mmWave) communication, and network heterogeneity \cite{andrews2014will, gupta2015survey}, among others. 
The mmWave spectrum will provide a very wide contiguous bandwidth in high frequency bands (e.g. 24 GHz, 28 GHz, or 38 GHz etc.).  
The heterogeneous network (HetNet), on the other hand, will increase the capacity by allowing load sharing between several kinds of cells or even radio access technologies (RAT) \cite{andreev2019future}. For example, in a HetNet, a sub-6 GHz macro-cell may co-exist with mmWave-enabled small cells. It is expected that the small cells will be densely deployed to assist the conventional macro-cell layer \cite{chen2019performance}. The roles of macro and small cell layers are not formally defined yet. 
However, a general consensus is that the macro-cell layer will cover a wider geographical region, giving services to a large number of cellular users. The macro-cell will use sub-6 GHz spectrum for wider coverage. 
The small cells, on the other hand, will provide services to relatively smaller but densely packed regions where traffic demands remain high such as inside stadiums, shopping malls etc. The small cells may support both sub-6 GHz and mmWave access link communication  \cite{kamel2016ultra}.   

\begin{figure}[t!]\centering
	\includegraphics[width=3.6 in, trim={3.0cm 4.0cm 3.3cm 3.6cm},clip = true]{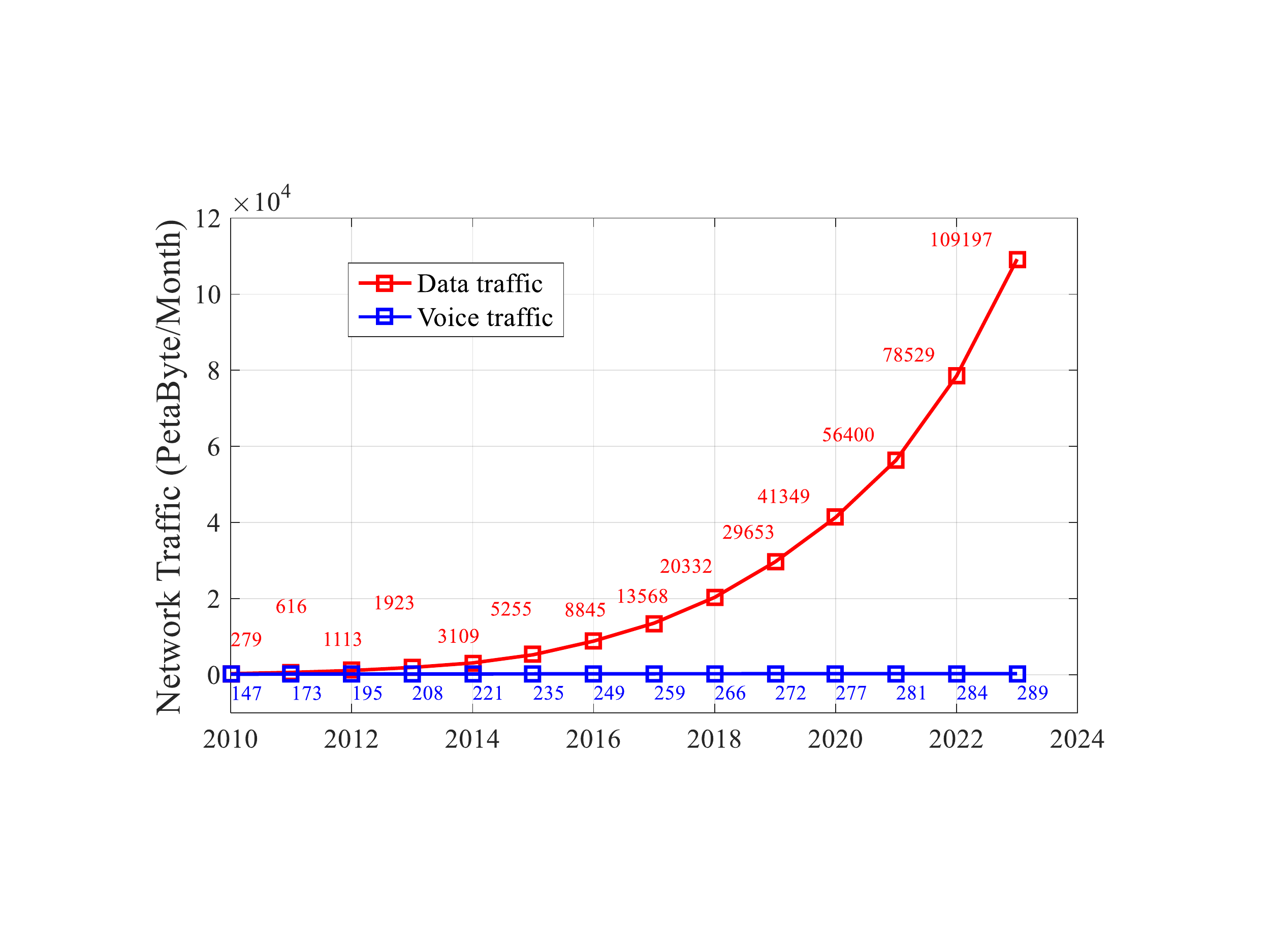}
	\caption{Extrapolated growth of voice and data traffic over the years [Source: Ericsson Mobility Report 2018]}	
	\label{figch1:data_vs_voice}
\end{figure} 

\begin{table*}	[!t]
	\renewcommand{\arraystretch}{1.3}
	\caption{Summary for Existing Surveys related to Moving Networks.} 
	\label{table_survey}
	\centering
	 
	\begin{tabular}{|p{.6cm} |p{.6cm} |p{6cm} |p{2cm}|p{5cm} |}
		\hline
		\textbf{Year} & \textbf{Paper} & \textbf{Survey Topic(s)} & \textbf{Relevant sections in this survey} & \textbf{Details in this Survey}\\ \hline
		2018 & \cite{chen2018development} & Mobile communication system for HSR, focusing on safety and control messages & \ref{subsec_HSR} & NA\\ \hline
		2018 & \cite{kim2018comprehensive} & mmWave enabled communication in railways &   \ref{sec_usecases}, \ref{sec_perf}, \ref{subsec_HSR} & Included sub-6 GHz and fiber-optic communication for railways, subways, and HSR\\ \hline
		2016 & \cite{jaffry2016making} & A short survey focusing on interference management in sub-6 GHz pertinent to moving networks &  \ref{subsec_interference} & Detailed discussion on interference management and resource allocation techniques in sub-6 GHz, mmWave, and fiber-optic transmissions. \\ \hline
		2015 & \cite{wang2015channel} & Channel measurement techniques for high speed trains &   \ref{subsec_HSR} & N/A\\ \hline
		2015 & \cite{moreno2015survey} & Radio communication for high speed rails with focus on safety and control message transmission &   \ref{subsec_HSR} & NA\\ \hline
		2010 & \cite{fokum2010survey} & Deployment of backhaul and fronthaul network for broadband in railway & \ref{subsec_HSR} & Detailed discussion about fiber-optic and mmWave backhaul for subways, trains, and high speed railways.\\
		\hline		
	\end{tabular}
\end{table*}

The motivation behind fixed small cells is to provide dedicated services to users in densely populated regions. The small cells are expected to increase the quality of service (QoS) for such cellular users by reducing the transmitter-to-receiver distances to lessen the distance-dependent large scale fading effects.  
The smaller geographical coverage of small cells will enable increase in network capacity by allowing aggressive frequency reuse through massive cell deployment \cite{kamel2016ultra}. Small cells will also be equipped with on-site cache to eliminate the redundant communication with the core network \cite{li2018contract}, thus reducing the content delivery time, along with freeing up the backhaul links \cite{yu2016backhaul,cheng2018localized}.  

\subsection{Rationale for Moving Networks}

Due to the reasons discussed above, increasing the density of fixed small cells has been considered as a mainstream solution by the research community to meet ever increasing data demands \cite{kamel2016ultra}. However, due to their smaller coverage area, fixed small cells are expected to serve stationary or low-speed cellular users (e.g. walking pedestrians or by-passers etc.) only. On the contrary, 5G networks will need to accommodate scenarios which are not yet encountered by legacy 4G/3G networks. 

\begin{figure}[t!]\centering	
	\includegraphics[width=3.5 in, trim={4.5cm 3.2cm 4.5cm 3.8cm},clip = true]{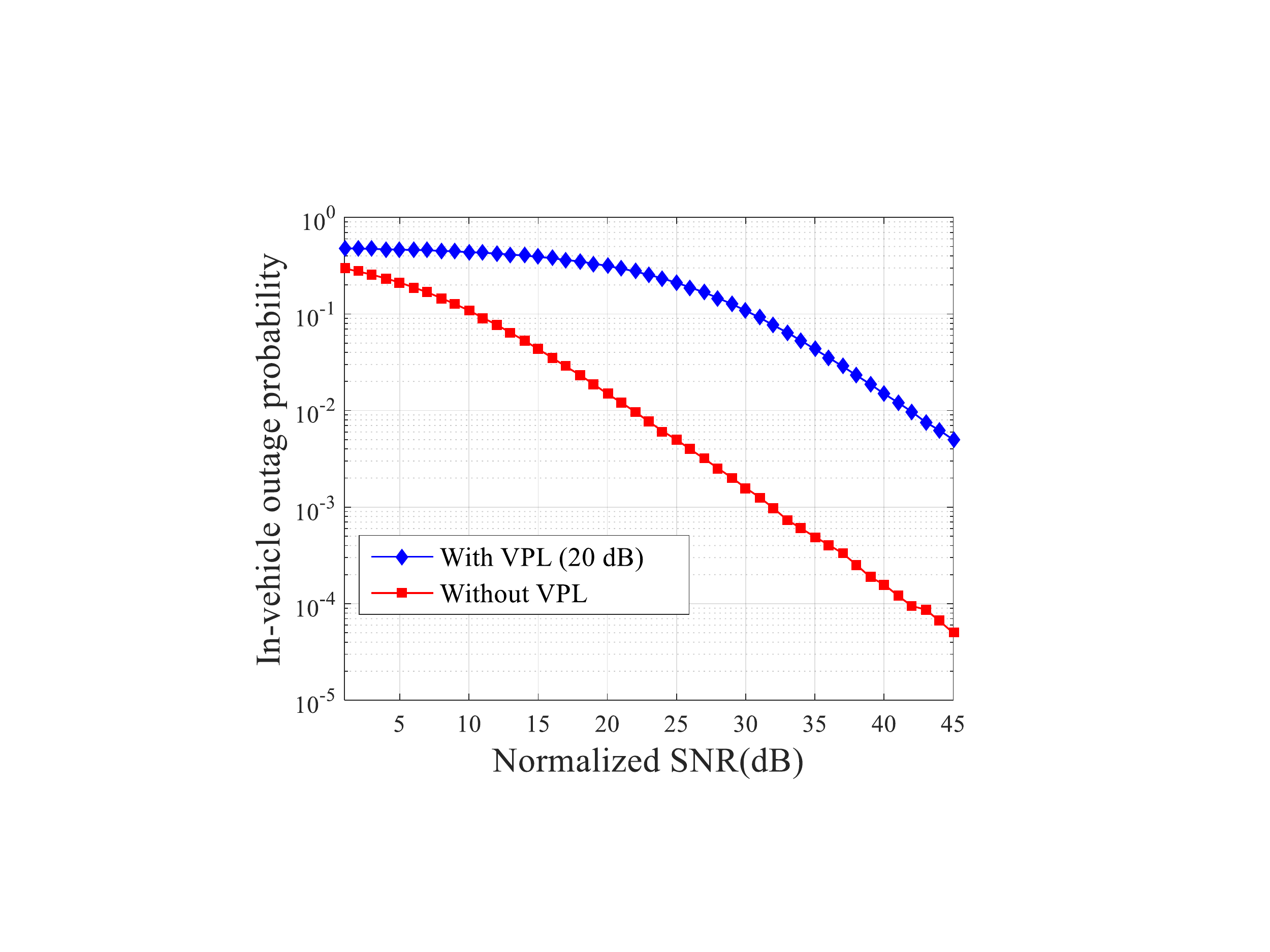}
	\caption{Effect of VPL on the outage probability.}
	\label{fig:VPL_effect_ch1}
\end{figure}

An example of such a scenario will be to provide uninterrupted cellular services to large number of commuters inside public transport vehicles such as trains, subways, buses etc. These travelers will make a large proportion of cellular users \cite{Ericsson2015}. These densely populated cellular users spend considerable amount of time in vehicles, such as, intercity travelers in high speed trains or buses etc. Currently, these commuters are served by the macro-cell base stations. Oftentimes, these commuters receive poor QoS due to Vehicular Penetration Losses (VPL) \cite{tanghe2008evaluation,tang2017coverage}. The VPL degrades the signal strength as radio waves  penetrate through vehicle's body as shown in Figure \ref{fig:VPL_effect_ch1}. In \cite{tanghe2008evaluation}, authors conducted experiments to demonstrate that VPL may degrade signal strength by as low as 25 dB for a sub-6 GHz frequency bands. For higher frequency bands, the signal degradation will get even worse. 
In addition to VPL, Doppler shifting due to high velocity of vehicles, such as, high speed railways (HSR) further worsens the signal quality for in-vehicle users \cite{rodriguez2015lte}. Furthermore, as we shall discuss later in greater detail, users inside public vehicle often exert excessive load on the network due to large group handovers. 

One solution to improve QoS for users inside public transport vehicles is to link commuters with the fixed small cells massively deployed throughout the transport route. However, this is not an efficient solution due to following reasons:
\begin{itemize}
\item  Fixed small cell base stations will be low-powered nodes and VPL will severely degrade the signal strength for in-vehicle users. 
\item  The handover rate will significantly increase due to shorter coverage of small cells \cite{lin2013towards}. 
\item The probability of link failure may increase with excessive handovers when vehicle moves frequently from the coverage of one cell to another. 
\end{itemize}

Due to the shortcomings of the above proposed solution, researchers have proposed to use dedicated small cells for public transport vehicles \cite{jaffry2016making}. Researchers have used different names, such as, moving small cells, moving relays, mobile relays, or mobile cells to refer to these dedicated cells for public transport vehicles. Henceforth we will only use the term mobile cells to generally address all the above mentioned terms. A collective network of mobile cells is called a moving network, which is the topic of our survey. 

In a mobile cell, the commuters will communicate only with an in-vehicle antenna, thus circumventing VPL and Doppler effects. A mobile cell will connect to the core network (or backbone network) through backhaul link. Neighboring mobile cells in a moving network will communicate with each other through sidehaul links. Researchers have demonstrated that the moving networks will increase the overall throughput and capacity of future cellular networks \cite{jaziri2016offloading,jaffry2019efficient}. 

Moving networks were primarily introduced to cater in-vehicle users \cite{jaffry2016making}. However, researchers have explored several new dimensions and use cases for moving networks, such as, mobile caching, adaptive networking etc. Despite the potential of moving networks, extensive research is required to explore its efficiency and usefulness before real-world deployment. For example, there are some existing research challenges, such as  interference issues, resource management problems etc. We shall discuss all use cases and research challenges at a greater length in this survey.

\begin{figure*}[t]\centering
	\includegraphics[width=\linewidth, trim={0cm 0cm 0cm 0.0cm},clip = true]{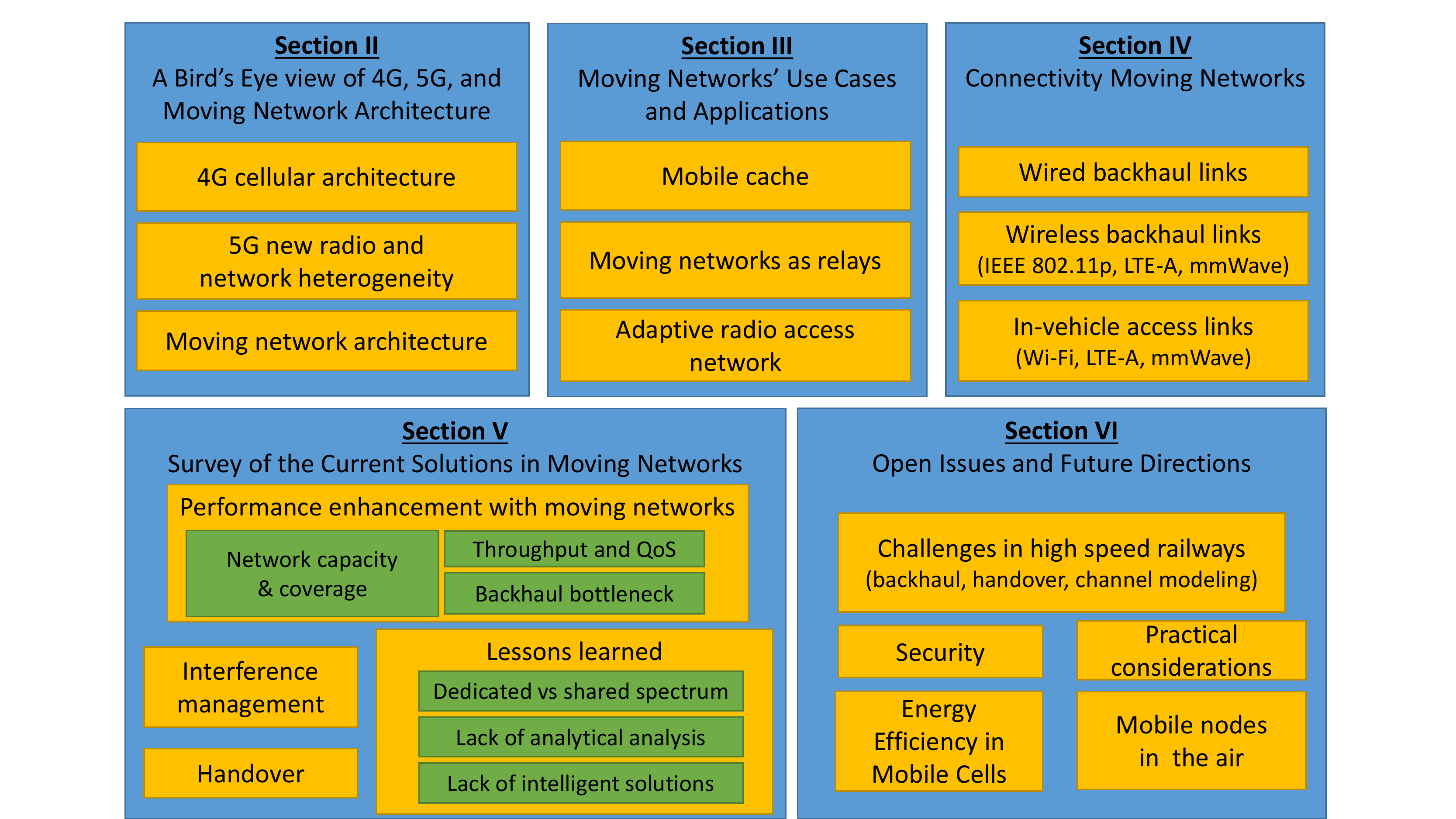}
	\caption{Organization of Survey.}
	\label{figch1:surveypresentation}
\end{figure*} 

\begin{figure}[t]
	\centering
	\includegraphics[width=3.5 in, trim={9.5cm 5.3cm 9.5cm 5cm},clip = true]{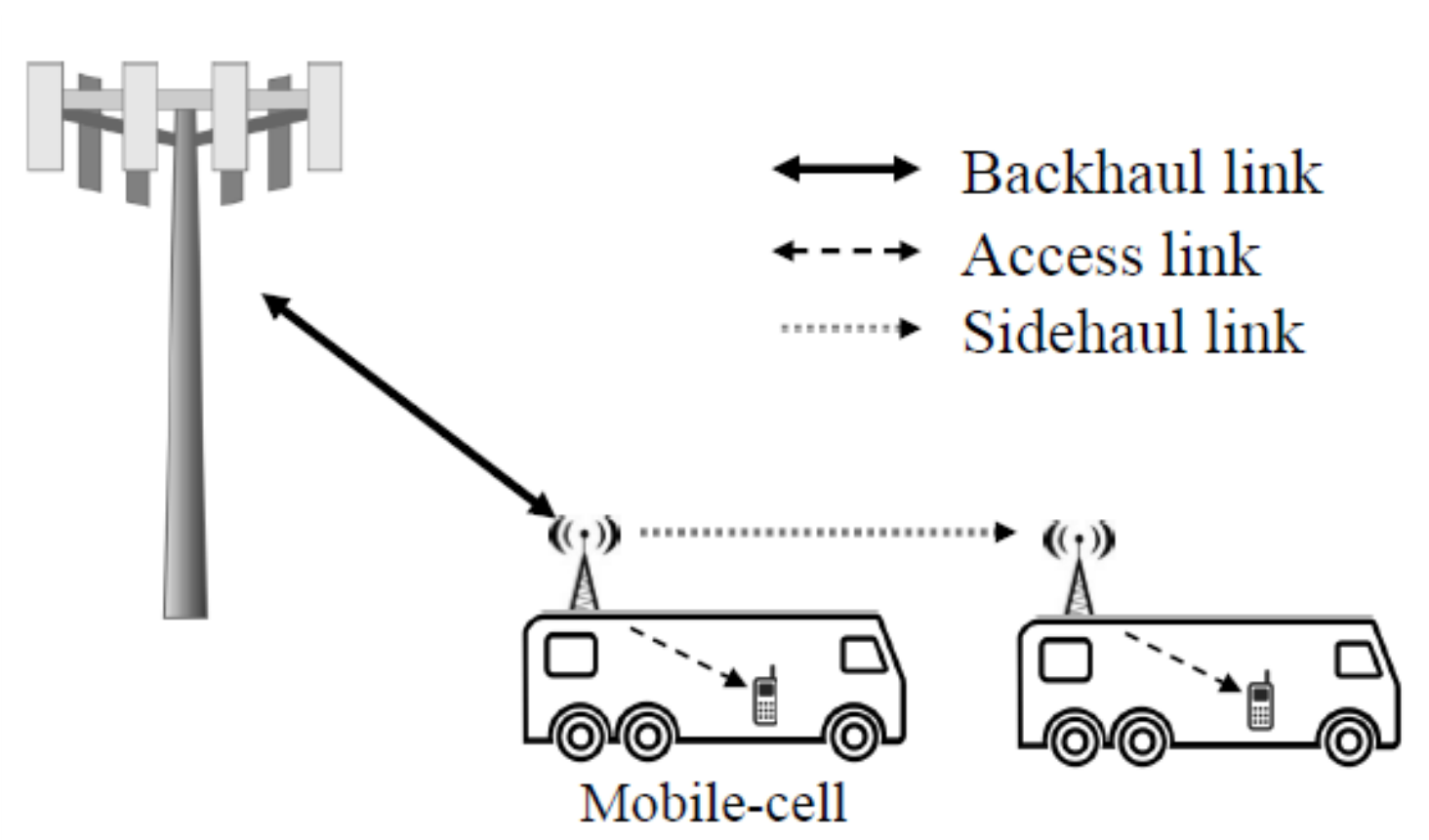}
	\caption{Mobile-cell's wireless links}
	\label{fig_mobile_cells}
\end{figure}

\subsection{Scope of this Survey}

This survey will present a comprehensive review on all the researches that have addressed any aspects of moving networks, including use cases and applications of a moving network. We will also discuss how mobile cells and relays will enhance different performance metrics in future cellular networks. Later on, we will present some of the challenges associated with the moving networks addressed by researchers. We will also put out some of the open questions that need further investigation. We will thoroughly examine all relevant research concerning moving networks that has taken place in at least last the 10 years. 

This survey does not cover or focus on the vehicular communication which takes place for emergency purposes or to share location information or state of vehicle's mobility. Such vehicular data exchange system, often known as intelligent transportation system (ITS), may include vehicle-to-vehicle (V2V), vehicle-to-infrastructure (V2I), or vehicle-to-everything (V2X) communication. A rich body of literature on vehicular communication and ITS already exists that can be referred to, for example, in \cite{wang2019survey,ahmed2018cooperative,siegel2017survey} etc. and  references therein. \linebreak

\subsubsection{\textbf{Survey Contribution}}

In Table \ref{table_survey}, we have listed all relevant surveys that may align with our work in this paper. Most of the researches such as \cite{fokum2010survey,moreno2015survey,chen2018development,kim2018comprehensive,  wang2015channel} have focused on one or few aspects of communication in railways while missing other transport vehicles such as buses, vans, cars etc. 
To the best of our knowledge, this is the first extensive survey that covers all aspects of moving networks including installations inside railways, subways, trains, buses, and other public transport. 

The main contribution of this survey are as follows:
  
\begin{itemize}
	\item We will give a detailed overview of moving networks, along with all the use cases discussed in the literature. 
	\item We will discuss the key challenges related to the integration of moving networks with the legacy static cellular layer of 5G and beyond. This survey will also discuss the solutions proposed by researchers in this domain.
	\item We will present how the researchers have examined performance enhancement of legacy static cellular networks with the integration of mobile cells. 
	\item We will also present some open unresolved research questions and exciting research directions related to moving networks.
\end{itemize}

\subsection{Survey Organization}
This survey is organized as follows. Section \ref{sec_main5Gfeature} gives a bird's eye view of 4G, 5G cellular architecture with focus on radio access technology. We also introduce moving network architecture in this section. Section \ref{sec_usecases} discusses use cases and applications of moving networks. Section \ref{sec_arch} contains a detailed discussion on the links associated with the moving networks such as backhaul and access links. In Section \ref{current_sol}, we discuss the current solutions related to moving networks issues as discussed by the literature, followed by a discussion on open issues and future directions in Section \ref{sec_challenges}. We conclude this paper in Section \ref{sec_conclude}. Table \ref{table_papers} at the end of the paper lists the important researches discussed in this survey. Figure \ref{figch1:surveypresentation} provides the survey organization chart.

We have divided this survey into several sections for better readability. However, any research may befit more than one sections. For example, a research that presents a novel use case of mobile cell by proposing a new architecture is discussed in all relevant sections i.e. Section \ref{sec_usecases} and Section \ref{sec_arch}. However, we will focus only on the relevant part of a given research in a respective section. 

\section{A Bird's Eye View of 4G, 5G, and Moving Network Architecture}
\label{sec_main5Gfeature}

In this section, we briefly discuss the architectural details of LTE-A (i.e., 4G) and 5G network architecture relevant to our discussion on moving networks. It is important to mention that we do not explore the in-depth details of any of these technologies and instead refer the readers to the existing literature  and text-books such as \cite{shafi20175g,molisch2012wireless, gupta2015survey, rost2016mobile, ahmadi20195g, elnashar2014design, dahlman20134g} etc. In the later part of this section, we will discuss the provisional architecture of moving networks.

\subsection{4G Cellular Architecture}

\begin{figure}[t]\centering
	\includegraphics[width=4 in, trim={5.5cm 2.0cm 2cm 2.0cm},clip = true]{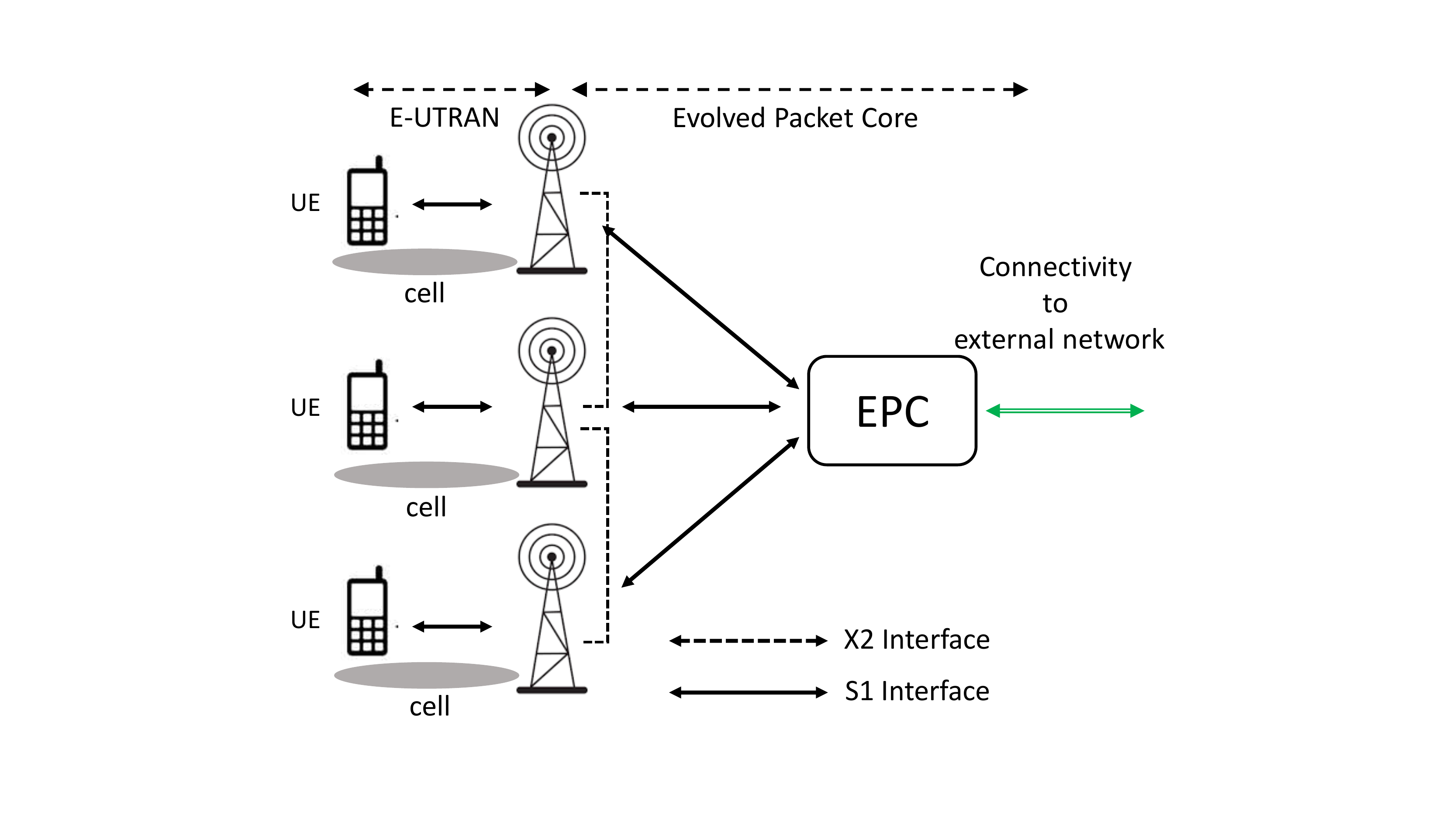}
	\caption{Architecture of Long Term Evolution - Advanced (LTE-A) network.}
	\label{figch1:LTEarch}
\end{figure}

The 4G network, also known as Long Term Evolution - Advanced (LTE-A), relies on the interconnected evolved NodeB (eNBs) \cite{elnashar2014design}. The Radio Access Network (RAN) is established between an eNB and the User Equipment (UE) over Uu interface. The radio interface is called enhanced Universal Mobile Telecommunications System (UMTS) RAN or E-UTRAN \cite{3GPP_EUTRAN_TR_R9} in LTE-A nomenclature. Neighboring eNBs are connected to each other via IP-based logical X2 interface in a peer-to-peer fashion. All eNBs are linked to the core network with the S1 interface. The core network in LTE-A is called Evolved Packet Core (EPC). Figure \ref{figch1:LTEarch} shows the LTE-A architecture.

The LTE-A RAN operates on sub-6 GHz frequency bands (e.g., 900/1800, 2100 MHz, and 2.4 GHz bands etc.). The earlier versions of LTE-A (i.e., LTE, also known as 3G cellular technology) offered a narrow bandwidth of up to 20 MHz whereas in LTE-A. The maximum bandwidth was expanded to 100 MHz in latest version of 4G. In 5G, the mmWave band will drastically expand the contiguous bandwidth to enhance capacity of networks. 

\subsection{5G Cellular Architecture} 

\begin{figure*}[t]\centering
	\includegraphics[width=0.8\linewidth, trim={2.0cm 2.5cm 2cm 2.0cm},clip = true]{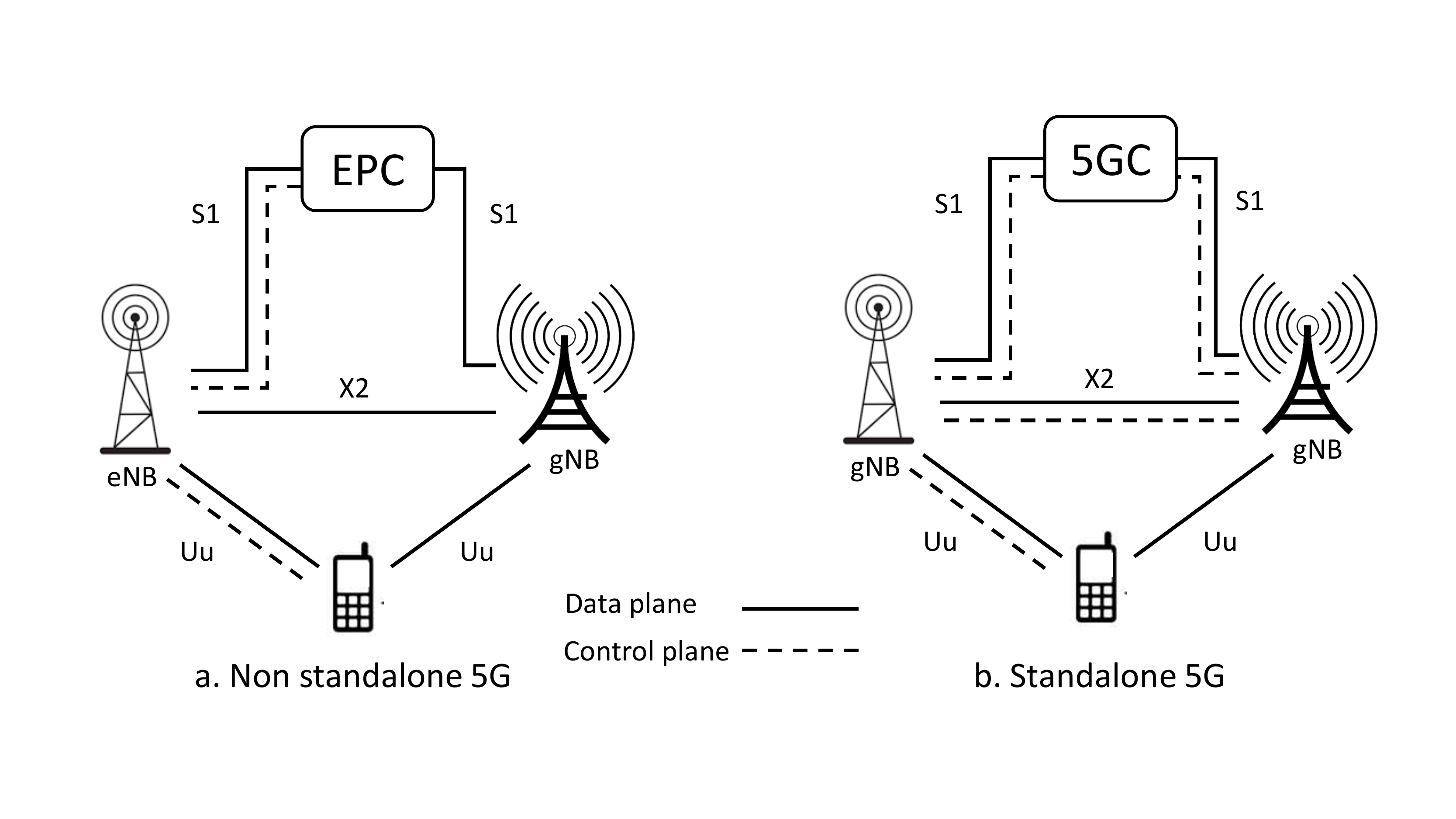}
	\caption{Architecture of 5G NR as defined by third generation partnership project specifications.}
	\label{fig:5Garch}
\end{figure*}

The global research on 5G radio access has been concluded with formal approval of 5G New Radio (5G NR) by Third Generation Partnership Project (3GPP) in 2017 \cite{3GPP_R15}. To expedite the deployment of 5G, a non-standalone architecture (NSA) was ratified which depends on LTE-A core (EPC). The 5G base station (also called gNB) can only provide data plane communication in NSA mode. Figure \ref{fig:5Garch}(a) shows NSA architecture where dotted lines refer to the control plane. Later on, a Standalone (SA) architecture will replace NSA with a separate 5G core (5GC, analogous to EPC in LTE-A) \cite{Ericson5GNR}. In the SA architecture, both control and data plane will be served by 5GC as shown in Figure \ref{fig:5Garch}(b).  

The 5G NR spectrum is divided into two Frequency Ranges (FRs). 
The FR-1 includes sub-6 GHz spectrum and may be used to provide coverage in wider geographical regions. 
The FR-2, which is often referred to as mmWave band, operates over 24 GHz (or above) frequency spectrum.  

The mmWave bands provide multi-gigabit transmissions \cite{lopez2019opportunities}. However, high freespace path loss and penetration losses for short wavelength mmWave signals will reduce the cell sizes \cite{akyildiz2018combating}. The use of highly directional antennas with Multiple-Input-Multiple-Output (MIMO) technology will compensate for inherent severe attenuation in mmWave \cite{dutta2019case}.  The overall architecture of future 5G networks is envisioned as a layer of small-cells (mmWave-enabled or otherwise) that may serve densely populated regions. Cells of different radio access technologies (mmWave and non-mmWave bands) and various sizes (macro, micro, femto, pico, etc.) are also envisioned to co-exist. This heterogeneous architecture is a key enabler in 5G networks \cite{ghosh20195g}.  

The network densification and load distribution between macro-cell and  several small cell layers are two pinnacle characteristic features of 5G HetNet \cite{andreev2019future}. It is important to note that fixed low-powered small cell base stations only provide improved QoS in static places such as offices, stadiums, shopping malls etc. However, the recent trend in public transport vehicles becoming new cellular hotspots, warrants new communication technologies. This is because of the recent advancements in connected and autonomous car technology and the increased in-vehicle user activities  \cite{jaffry2018effective,julsrud2017smartphones,Deloite_survey2017,Ericsson2015}. Therefore, moving networks will become an essential part of 5G and beyond network architecture.

\subsection{Moving Network Architecture}
\begin{figure}[t]\centering
    \includegraphics[width=\linewidth, trim={4.0cm 4cm 4cm 3.8cm},clip = true]{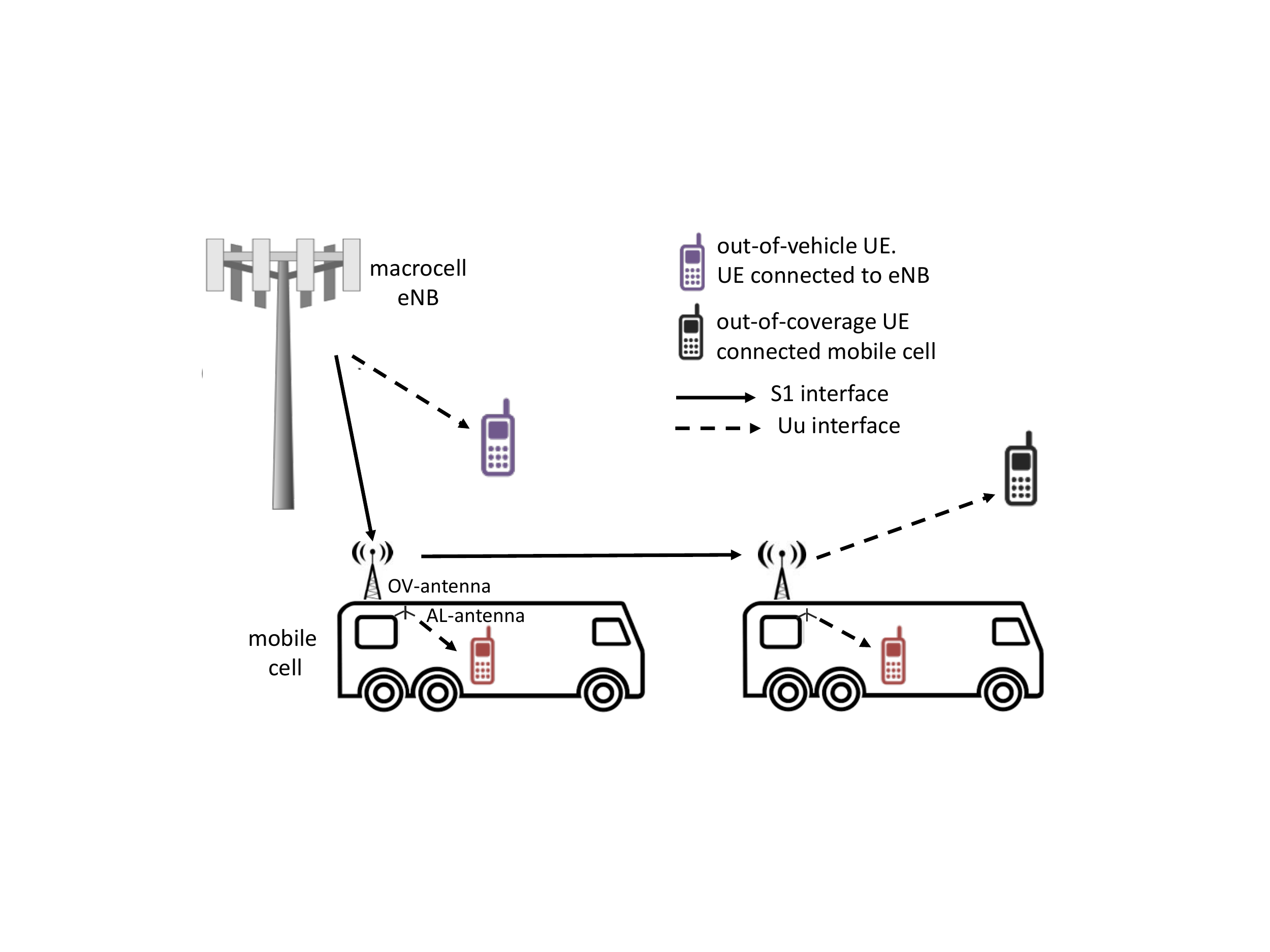}
    \caption{Mobile cell interface}
    \label{fig:mobilecell_interface}
\end{figure}

Moving network is formed through the interconnection of mobile cells where mobile cells are installed inside public transport vehicles such as trains or buses as shown in Figure  \ref{fig:mobilecell_interface}. The aim of mobile cells is to improve the QoS for in-vehicle users. To date, research community has explored and identified several use cases for mobile cells.   

From the design perspective, there are multiple antennas on-board a mobile cell as shown in Figure \ref{fig:mobilecell_interface}. The Access Link antenna (AL-antenna) is responsible for enabling communicate between commuting users and the transport entity (bus or train). Whereas the out-of-vehicle antenna (OV-antenna) connects a mobile cell to the core network, to the neighboring mobile cells, or to the out-of-vehicle users (e.g. pedestrians etc.)

The AL-antenna is installed inside a vehicle and may establish a \textit{near} Line-Of-Sight (LOS) link with the in-vehicle devices. The proximity of AL-antenna to the on-board cellular users is used to mitigate the large scale fading effects, and hence improving the signal quality at the receiving end. Moreover, the AL-antenna serves as the anchor point for all the communication that takes place inside a mobile cell. The access link uses {\it Uu} interface to communicate with the commuting users. Additionally, the AL-antenna may take the role of either eNB or gNB depending on NSA- or SA-based network deployment.

The OV-antenna ensures the connectivity of mobile cell with the core network over backhaul link and with the neighbors over sidehaul links. A backhaul link can be both wired or wireless connection depending on the specified use-case. The backhaul link communicates over X2 interface with the nearby macro-cell base station. 
The sidehaul links establish connectivity with the neighboring mobile cells or the out-of-vehicle cellular users \cite{shin2016sidehaul}. In the later sections we will show how cellular users outside the vehicle can benefit from moving networks' sidehaul links. The sidehaul communication also takes place over X2 interface. The OV-antenna and AL-antenna are linked together using Common Communication Unit (CCU) which also controls resource and power allocation for access, backhaul, and sidehaul links \cite{jaffry2019interference}. It is important to mention that the architecture of moving networks is not yet standardized and a detailed discussion is presented in Section \ref{sec_arch}.

Other elements that may compliment a moving network can be Road-Side Units (RSU) \cite{Hussain2017,Kim2017,jolfaei2019secure}. RSU can be either installed in the optimized locations that are close to mobile cell's route or there could be mobile RSUs that could increase the connectivity \cite{Hussain2014,Hussain2017} and these RSUs can be dedicated fixed small cells or fixed relays. 
In the literature, several operational use-cases of mobile cells have been investigated which we have classified into three main categories, i.e., mobile cache, mobile relay, and adaptive radio access connectivity. It is worth noting that these use-cases may overlap with each other.    
In the following section, we discuss these use-cases. 

\section{Moving Networks' Use-Cases and Applications}
\label{sec_usecases}

In this section, we discuss different use-cases of moving networks.  The use-cases can be abstractly divided into three categories. The first category is the role of moving network as mobile cache, second category is mobile relay and third category is the provision of adaptive radio access connectivity through moving network. These categories are further elaborated below.

\subsection{Mobile Cache}
\label{subsec_mobilecache}
The mobile data traffic has exponentially soared in the last few years due to popularity of applications such as video streaming, online gaming, video calling, or plain web surfing \cite{ciscovirtual_report2019,cerwall2018ericsson} (See Fig. \ref{figch1:data_vs_voice}). 
Furthermore, the trend of using Internet in the public transport by the commuters has also exponentially increased. 
Therefore, the research community and service providers are making huge efforts to install on-board cache in the future public transport vehicles.
These mobile caches will save significant network bandwidth by reducing the number of frequent data requests to the core network \cite{shah2018moving,shin2016sidehaul,jiang2015spatio}. Jiang et al. \cite{jiang2015spatio} demonstrated that commuters traveling inside public buses often make similar data requests. The authors analyzed the data collected from more than 20 inter-city buses commuting on six different routes in Sweden. With Spatio-temporal examination of the contents, the authors argued that the installation of an in-vehicle cache could save 20\% of daily bandwidth. However, one of the critical challenges with the mobile caches is to develop an efficient content update mechanism in fast moving vehicles.  

In \cite{kwon2015radio}, Kwon et al. presented a proactive content update mechanism for cache within mobile cells. They used evolved Multimedia Broadcast and Multicast Services (eMBMS) of LTE-A to update cache.  The decision to use broadcast or multicast transmission was made using content popularity distribution (such as ZipF distribution \cite{zipf1929relative}) at the dedicated servers installed with the macro-cell base stations. The authors used orthogonal spectral bands for macro-cell's and mobile cell's access links. Furthermore, macro-cell transmissions and reception used 2.0 GHz band, and access link inside mobile cells used 3.5 GHz band. For the backhaul links, the mobile cell shared spectrum with macro-cell using in-band full-duplex mode. Additionally, the authors also investigated the content sharing between neighboring mobile cells based on popularity index using 2.0 GHz band. They reported that the network's throughput increased proportionally with the increase in contents' popularity. 
Another factor that contributed to the increase in throughput, was the density of mobile cells. With more number of mobile cells in a given area, neighbors share contents more frequently with each other, relieving the macro-cell base station. This sidehaul content sharing further spared frequency spectrum for macro-cell legacy access link communication.  

\subsection{Moving Networks as Relays} 
\label{subsec_relay}
As a relay, a mobile cell can carry information from the macro-cell base station to the in-vehicle users with a dual hop, circumventing the VPL. The existing researches have demonstrated significant increase in QoS for users inside the ``mobile relays'', despite the addition of a single hop \cite{shah2018moving,sui2012performance,iturralde2018performance}. From the standardization standpoint, 3GPP ratified specifications for moving relays in Release-12 \cite{3GPP_spec_MobileRelay}.   

In \cite{iturralde2018performance}, The authors analyzed QoS enhancement for commuting users using mobile relays in the railways. The authors compared in-vehicle communication when users were served with mobile relays versus when they were directly connected to a nearby fixed base station. The authors focused on HTTP and Voice over IP (VoIP) service delivery because 4G uses VoIP for telephony services as well. Due to the relaying process, additional headers were introduced in the backhaul transmission packets.  
The authors reported that eliminating the VPL using mobile relay improved QoS for passengers; however, a small cost of additional headers was incurred in the message (40 Bytes). The authors observed that a standard website has a size of 150 KB up to 8 MB which makes the additional header size negligible. The authors also reported  that the median webpage's load time reduced from 7 sec to 2.5 sec for users inside mobile relays. The maximum load time was reduced by 30 seconds. The authors also reported that the throughput (measured in Mbps) for direct and relayed service was similar when the train was located near the base station. However, when the train was at a distant location from the base station, mobile relay improved the throughput from 0.1 Mbps to 0.25 Mbps (direct vs relayed services). Due to a single hop in the mobile relay as compared to direct transmission, the VoIP latency slightly increased from 25 ms to 37 ms. However, authors argued that latency is still much lower than the International Telecommunication Union (ITU) recommended period of 150 ms. 

There are also a handful of mechanisms in the literature that investigated the use of mobile cells to increase the coverage of out-of-vehicle cellular users though cooperative relaying \cite{li2012cooperative,khan2017outage,feng2017vehicle}. 
For example, the authors in \cite{feng2017vehicle} demonstrated that mobile cell assisted communication in dense metropolitan areas can improve network connectivity for out-of-vehicle users (through cooperative relaying). In \cite{tang2017coverage}, the authors investigated Coordinated Multi-Point (CoMP) joint transmission between mobile relay and macro-cell base station to serve out-of-vehicle users. The authors exploited the fact that mobile cells will be densely distributed in future urban environment. They also reported a drastic decrease in outage (equivalent to increase in coverage) due to joint transmission from macro-cell and mobile relay. In their network setup, cellular users were jointly connected to the macro-cell base station and the nearby mobile relay based on certain interference threshold condition. Furthermore, the authors found a strong relation between the coverage probability and density of mobile cells. Finally, the authors reported that as the number of mobile cells grew above a certain limit in an area, coverage probability decreased due to high interference from the nearby transmitting mobile cells. 

\begin{figure}[t]
	\centering
	\includegraphics[width=3.5 in, trim={3.8cm 5cm 1.0cm 6cm},clip = true]{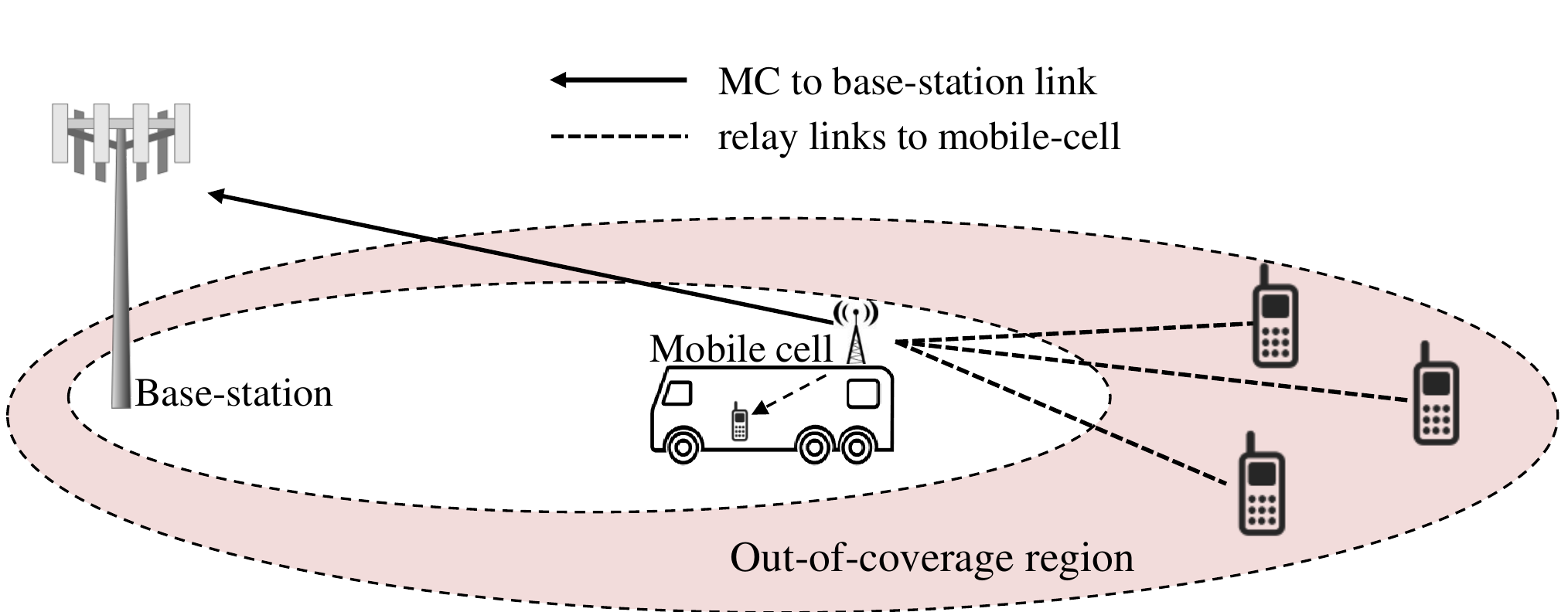}
	\caption{Mobile cell assisting communication in out-of-coverage region as presented by Shin et al. [31]}
	\label{fig_MC_out_of_cov}
\end{figure}

In \cite{MShinPublicSafety}, Shin et al. used mobile cells sidehaul links to extend the cellular coverage to the unconnected regions as shown in Figure \ref{fig_MC_out_of_cov}. It is important to note that sidehaul communication can share the cellular uplink spectrum as specified by 3GPP or can use the dedicated 700 MHz spectrum for public safety application \cite{3GPPpublicsafety,shah2018moving}.  Similarly, Shah et al. \cite{shah2018moving} also investigated the dual role of a mobile cell as a relay and as a cache.  
Since a vehicle does not have limitation of power on board, a mobile relay may connect to the core network with stronger backhaul link as shown in Figure \ref{fig_MC_out_of_cov}. However, the backhaul links are the main bottleneck that may restrict and adversely affect system throughput. Hence, in \cite{shah2018moving} Shah et al. proposed a mechanism to increase the bandwidth at the backhaul by using orthogonal resources. 

In the next subsection, we discuss another use case, i.e., adaptive Radio Access Network (RAN) that stemmed from cooperative relaying.

\subsection{Adaptive Radio Access Network}
\label{subsec_adaptiveRAN}
There are several similarities between moving networks' role as cooperative relays and the adaptive RAN, such as, offering services to the out-of-vehicle macro-cell users. However, we believe that the ability of mobile cells to provide adaptive RAN needs a separate discussion. As an adaptive RAN, a mobile cell can provide temporary services in case of spontaneous cellular demands. There can be two possible scenarios that require adaptive RAN services. 

\begin{figure}[t]	\centering
	\includegraphics[width=4 in, trim={3.5cm 2.2cm 2.0cm 3cm},clip = true]{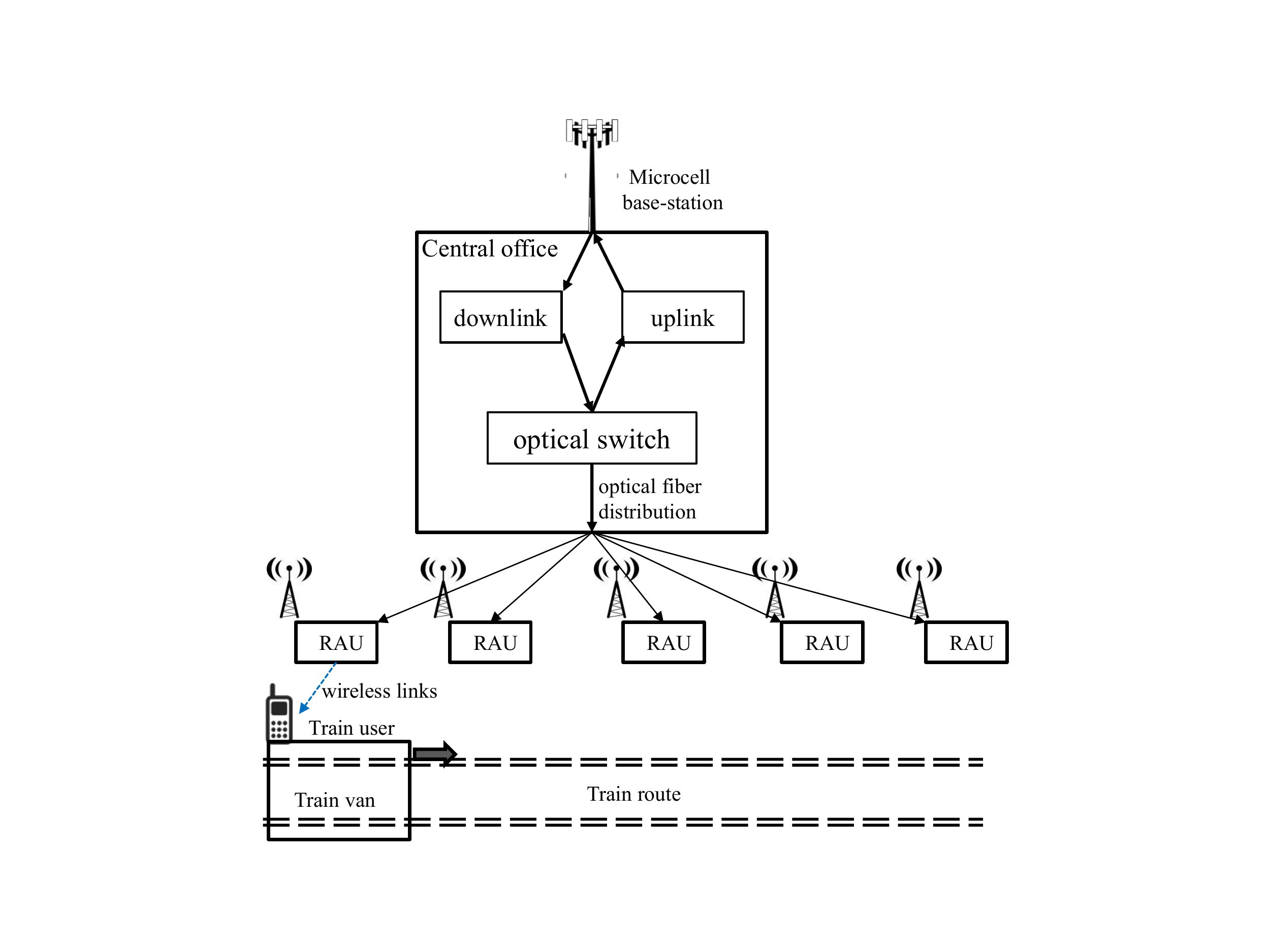}
	\caption{Radio-over-fiber technology as used in \cite{hsueh2011novel}}
	\label{fig:radio_over_fiber}
\end{figure}


In the first scenario, cellular operators may need to respond quickly in case of disasters or natural calamity. It is because, the network infrastructure is usually destroyed in such scenarios and need quick repair. Services providers are required to quickly establish a basic network connectivity. Recently, the 3GPP has put great emphasis on developing public safety standards for emergency use-cases  \cite{favraud2016toward}. In the literature, there are schemes that use mobile cells for public safety applications \cite{MShinPublicSafety,favraud2016toward}. Mobile cells may communicate with the out-of-vehicle users over sidehaul links and with the core network over wireless backhaul links. However, sidehaul link in this case will comprise of Uu interface, unlike X2 interface required for inter-mobile-cell communication. Similarly, in \cite{jaffry2018mobile}, the authors proposed the use of mobile cells to provide wireless services to the mobile black spot regions, i.e., regions where cellular coverage is very poor due to natural geography or terrain. Hence, like the example above, mobile cells may drive to these regions to provide extended coverage. A main challenge for mobile cells in public safety environment will be to quickly discover devices in the proximity  \cite{jaffry2019d2d}. 

The second scenario in which operators may need to temporarily augment their network capacity can be large-scale public gathering events such as sports weeks or festivals etc. \cite{andreev2019future,mirahsan2015hethetnets}. Cellular operators may need to temporarily expand their network capacity to accommodate a short and sudden surge in the traffic demands. Installation of dedicated fixed small cell infrastructure will be a costly solution. As a cost effective solution, mobile cells may quickly drive to the event venue to satisfy user traffic demands. These mobile adaptive RAN can return back to the stations or get deployed in other regions (in an on-demand basis) once their services are no longer needed.

In this regard, the authors in \cite{mohammadnia2019mobile} reported adaptive network densification using moving networks as a cost effective solution to share network load in excessively dense cellular regions. The authors argued that fixed small cell deployment is a costly, and oftentimes inefficient CAPital EXpenditure (CAPEX) for the network operators. Hence, installation of mobile cells inside vehicles (both public and private) can reduce the cost of network deployment. The authors also used an earlier study that reported that human mobility patterns in urban environment are often predictable. For example, the business districts in any urban area have peak data demand during office hours. On the other hand, the data traffic for the residential areas remains high during the evening hours or during weekends. The authors claim that installing large amount of fixed small cells in either environment will mean under utilization of the resources and capital investment. Instead, the installation of base stations to mobile vehicles that may carry users from residential to the business district and back, can provide cost effective solution. The authors in \cite{mohammadnia2019mobile} also performed evaluations using real-world data from the busy district in Milan, Italy. It was reported that a maximum increase of 150\% in throughput was achieved as compared to legacy fixed small cell infrastructure (in the absence of fairness in use). Considering the user resource allocation fairness, the gain is still 120\% higher than that of fixed small cell settings. 

\begin{figure*}[t]\centering
	\includegraphics[width=5 in, trim={2.5cm 1.8cm 2.0cm 1.5cm},clip = true]{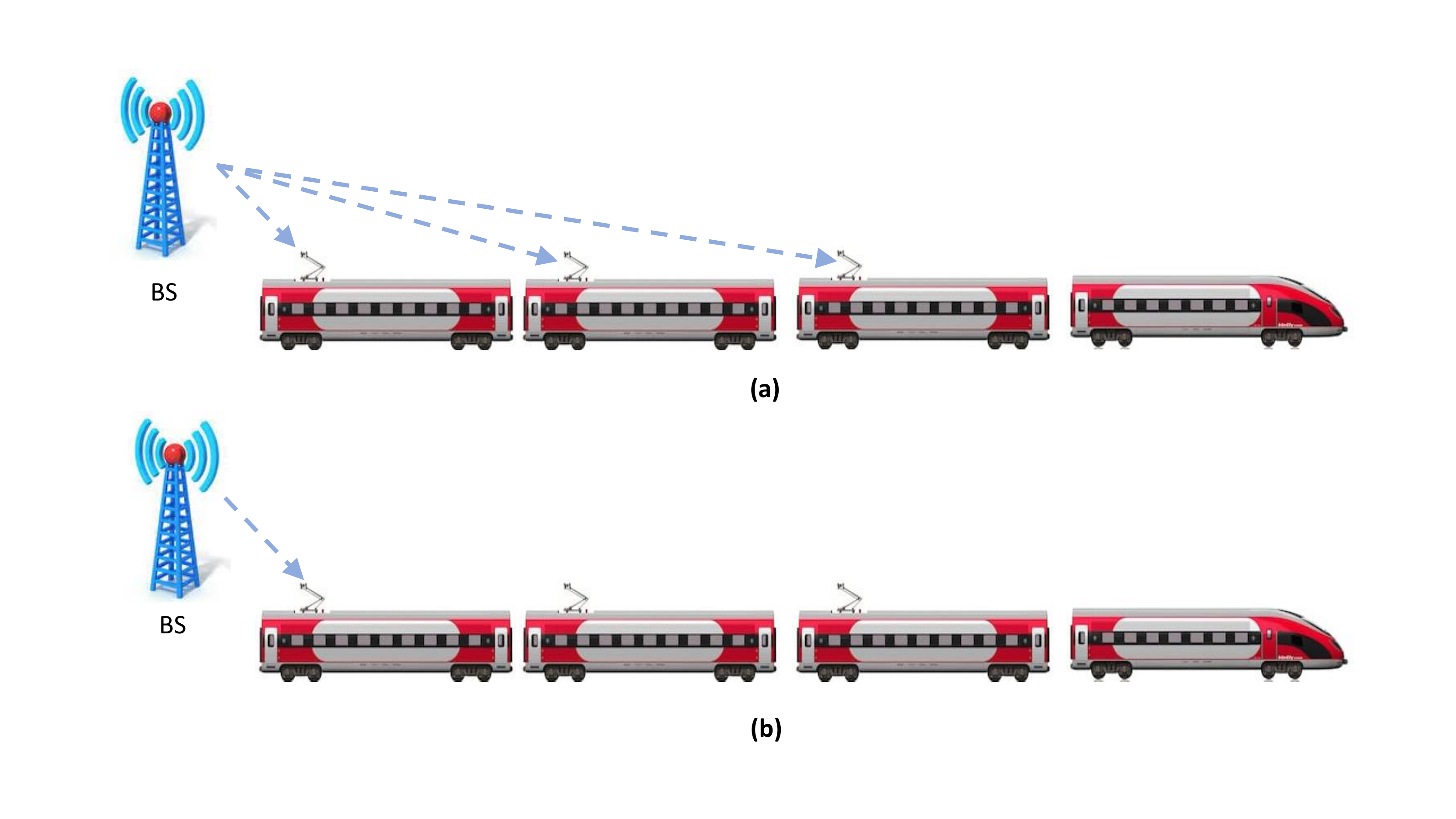}
	\caption{Different backhaul architectures for trains. Fig. (a) demonstrates the case when each van of the train is an individual mobile cell with a separate backhaul link to the macro-cell base station. Fig. (b) shows that the whole train is a single mobile cell. Each van is connected through Internet network within a mobile cell.}
	\label{fig_van2010}
\end{figure*}

In another work \cite{jaziri2016offloading}, the authors reported the similar trends. However, the authors also warned that the inefficient deployment of mobile cells would cause severe interference to static cellular layers. They performed experiments to demonstrate that an unplanned and random mobility decreases the overall network performance. In particular, when a mobile cell moves towards out-of-vehicle users that share the same sub-channels as in-vehicle users, devices sharing the same resources experienced severe interference. The high interference between in-vehicle and out-of-vehicle users consequently reduced the system throughput. On the contrary, the planned mobility of mobile cells significantly improved the QoS for all users, particularly at the cell-edges.

Next, we discuss the architectural level details of moving network in the light of the existing literature. 

\section{Connectivity in Moving Networks} 
\label{sec_arch}

The architectural challenges of a moving network are very  different than that of any fixed network. For example, unlike fixed small cells, moving network's mobility makes backhaul connectivity more challenging. In most practical scenarios, wired backhaul that can provide large bandwidth for fixed small cells, is not an option for the mobile cells. For example, buses operating in urban environment cannot have wired backhaul links. On the other hand, in the literature, some solutions use optical fiber based backhaul links for public transport vehicles following a fixed track, such as, trains or subways  \cite{nakayama2017optically,nakayama2018optically,hsueh2011novel}. In this section, we discuss the architecture level details of moving networks. 

\subsection{Wired Backhaul Links}
Recently, Optically Backhauled Moving Network (OBMN) was proposed to use in trains and subways \cite{nakayama2018optically,nakayama2017optically,hsueh2011novel}. The authors in \cite{nakayama2018optically, nakayama2017optically} reported that OBMN can provides very high bandwidth for backhaul links that enable stable connectivity for the on-board users. Similarly, in another work \cite{nakayama2018optically}, an OBMN-supported mobile cell was connected to the core network through Autonomous Base Stations (ABSs). These ABSs were deployed at multiple stops (e.g., railway stations) in the path of a moving train. The ABSs were also connected to the core network through wired broadband links. As a backhaul link for moving trains, a fiber optic reel was attached from the mobile cell's roof to the ABS. Furthermore, on-board users were connected to the in-vehicle antenna over wireless access links. From the afore-mentioned experimental environment, the authors reported a drastic decrease in the load of the core network with OBMN implementation. The authors also reported a sharp decline in the number of handovers using OBMN architecture because the fast moving in-vehicle users were decoupled from the macro-cell base station.

In \cite{hsueh2011novel}, the authors proposed a hybrid optical and wireless system based on Radio-over-Fiber (RoF) technology to provide reliable backhaul for broadband services to the railway commuters. As shown in Figure \ref{fig:radio_over_fiber}, the macro-cell base station communicates wirelessly with the central office for railway networks. At the central office, radio signals are converted into optical transmissions and distributed to the Remote Access point Units (RAUs). The RAUs are deployed at fixed locations enroute the railway track at regular intervals. Each RAU covers a small region and provides services to the train in the areas under its coverage. The authors also reported that the number of handovers decreased drastically as core network was decoupled from the end user inside fast-moving train. 

In essence, optical backhaul provides a very large bandwidth and high speed links between the mobile cell and central core. However, laying fiber optic for backhaul requires excessive CAPEX \cite{jaber20165g}. Rolling out optical fiber to cover the entire subway or railway track may also incur additional OPerational EXpenditures (OPEX) required for maintenance. A viable alternative is then to consider wireless backhaul for moving networks. 

\subsection{Wireless Backhaul Links}
The wireless backhaul has low cost of deployment (i.e., CAPEX) and is more suitable for outdoor urban environments. Hence, majority of researchers have considered wireless backhaul for moving networks \cite{sui2014deployment,zafar2019resource,sui2013moving,chae2013seamless,jaffry2019interference,jangsher2017backhaul,jaffry2018shared,jaffry2018effective,tang2020meta} etc. These researches discuss two competing technologies (IEEE 802.11p and cellular backhaul) to connect mobile cells to the backbone network.  
The first technology, IEEE 802.11p \cite{Jiang2008} is a variant of IEEE 802.11 standard and was originally designated for Dedicated Short Rage Communication (DSRC) for vehicular communication \cite{vinel20123gpp, cao2020accurate}. 

In \cite{mollah2016novel}, the authors used 802.11p standard to establish a wireless backhaul link between a mobile cell and the Road-Side Units (RSU). The RSUs cache and broadcast popular web contents (such as audio, images, and videos etc.) to the mobile cells based on a popularity index. The authors reported a trade-off between the system throughput and fairness of service in conventional scheduling mechanism as 802.11p standard offers only limited bandwidth. In \cite{khairnar2013performance}, the authors reported that the inherent Carrier Sense Multiple Access (CSMA) medium access (MAC) protocols of 802.11p could drop $80\%$ of vehicular data packets due to channel congestion. Hence, we can conclude that IEEE 802.11p may be a suitable technology for short emergency messages in Intelligent Transportation System. However, it does not befit the backhaul of mobile cells that needs very large bandwidth and greater transmission ranges. Hence, the research community have focused on a competing cellular backhaul. 

\begin{table*}	[!t]
	\renewcommand{\arraystretch}{1.3}
	\caption{Pros and Cons of Backhaul links.} 
	\label{table_backhaul}
	\centering	 
	\begin{tabular}{|l | l|  l| }
		\hline
		\textbf{Characteristic feature} &  \textbf{Optical Backhaul} & \textbf{Wireless Backhaul}\\
		\hline
		Serving Technology &Optical-fiber transmission&LTE-A or mmWave transmission \\ \hline
		Bandwidth support & Very large (multi-gigabits per second) & Small to medium (megabits to few-gigabits per second)\\\hline
		Interference to neighboring links & None & Possible\\		\hline
		Mobility support &Only deterministic& Deterministic and random\\\hline
		Deployment cost &High& Low\\\hline
		Maintenance cost &High& Low\\\hline
		Handovers &None&High \\\hline
		Connectivity to cellular network &In-direct (multi-hop) links&Direct\\\hline
		Provision of in-vehicle data services &Yes&Yes\\\hline
		Provision of in-vehicle cellular services &In-direct&Yes\\\hline
	\end{tabular}
\end{table*}

Sui et al. \cite{sui2014deployment} demonstrated the capacity increment of backhaul links using multi-antenna system with the integration of techniques such as maximum ratio combining and interference rejection combining. Similarly, in \cite{van2010providing}, the authors proposed a cellular Coordinated and Cooperative Relay System (CCRS) for the backhaul of mobile cells within trains. The authors introduced two backhaul designs as shown in Figure \ref{fig_van2010}, i.e., (i) inter-connected multi-mode CCRS (CCRS-1), and (ii) Distributed Antenna System (DAS) based CCRS (CCRS-2). In CCRS-1, as shown in Figure \ref{fig_van2010}, a train with \textit{X} coaches had \textit{X} individual mobile cells, each with a separate backhaul link to a donor base station. The mobile cells can communicate with each other over the cooperative interface called crX2 (similar to X2 interface in LTE/LTE-A). The crX2 interface can be wired or wireless logical links between neighboring mobile cells. On the other hand, the CCRS-2 architecture has a single backhaul (See Fig. \ref{fig_van2010}(b)) with multiple remote antenna systems installed throughout the train's coaches. The CCRS-2 design reduces the number of resources used for backhaul links. In \cite{van2010providing}, all Remote Antenna Units (RAUs) were linked together. Note that in both designs, each coach has separate in-vehicle AL-antenna, but CCRS-2 needs more coordination between individual coaches to synchronize transmission and reception from the backhaul. 

In \cite{chae2013novel}, Chae et al. used wireless transmission for high speed railways and proposed LTE-A Coordinated Multi-Point (CoMP) transmission for  mobile cell's backhaul links. The authors proposed a handover mechanism for the train coaches as they move from the coverage of a resident macro-cell to the neighboring macro-cell. Train's individual coaches were installed with dedicated indoor access points. The passage of train from the coverage of its serving cell to the neighboring cell triggered the handover mechanism to shift connectivity to the target cell based on signal strength. The authors showed that as the front compartment underwent a shift in handover, the last compartment remained connected to the serving cell base station. The train was also installed with a Central Controller (CC) that connected all the individual (in-vehicle) access points. Hence during the handover transition phase, the first compartment still received services from the serving base station connected to the last compartment of the train. This mechanism enabled smooth handover shifting in fast moving trains with large number of compartments. The authors reported a 40\% reduction in outage probability of the on-board users. 

In \cite{yasuda2015study}, the authors conducted experiments to investigate the performance after installing fixed small cell base stations near railway lines to provide wireless backhaul for trains. The authors reported a drastic improvement in average system throughput due to the installation of nearby small cell base stations. The throughput exceeded 30 Gbps when the mobile cells moved close to the fixed base stations. The minimum throughput remained close to 5 Gbps even when mobile cells moved away from ground base stations. The authors also reported that direct communication between mobile cell and macro-cell base station could only achieve an average throughput of 1 Gbps. Researchers used massive Multiple Input Multiple Output (massive-MIMO) antennas for backhaul. The authors found that mobile cells inside trains could reduce control signal by 66\% as compared to conventional scheme i.e., a direct link between mobile cell and macro-cell base station. One reason to achieve the improved results is the separation of control and user plane that authors considered during their experiment. It was also observed that transport vehicle design and passenger movement models affect the control signaling.
For example, the number of entrance/exit doors inside the train made a huge impact on signaling and network load. With larger number of entry/exit points, the instantaneous control signaling was also increased. In particular, a peak in the control signal transmission was observed as cellular users moved in or out of the train. Nevertheless, the authors reported a significant reduction in control signaling due to the decoupling of commuting users from the macro-cell base station.

In another work \cite{iftikhar2019resource}, the authors proposed to install fixed Radio Resource Head (RRH) units near train's track to provide mmWave wireless backhaul to the trains. The fixed RRHs were connected to the core network through microwave wireless links. Using heuristic models, the authors designed a maximum coverage problem to ensure that RRH was connected to maximum number of mobile cells. In a tightly defined system model, the authors reported the effectiveness of such a model. However, the fast moving train changes connection to RRH as it moves rapidly. For example, at a moving speed of 300 km/hour, a train could connect to more than 16 RRH. This phenomenon incurs a crippling amount of signaling overhead required for handover which was not considered by the authors. In Section \ref{subsec_HSR}, we will show the existing works on solving the handover problems in mmWave-enabled backhaul for fast moving trains. 

In Table \ref{table_backhaul}, we provide a comparison of the competing backhaul technologies discussed in this section. Next, we discuss the different access link technologies proposed for access link communication inside mobile cells.

\subsection{In-vehicle Access Link}

In a mobile cell, the commuting cellular users connect to an in-vehicle AL-antenna as shown in Figure \ref{fig:jaffrymobilecell}. The commuters will be decoupled from the core network. This decoupling will enable high QoS for the users by circumventing VPLs and reducing the transmitter-to-receiver distance. However, in order to provide high QoS for commuting users, the mobile cells' access link should not interfere with the out-of-vehicle communication. 

\begin{table*}	
	\renewcommand{\arraystretch}{1.3}
	\caption{Pros and Cons of WiFi vs Cellular Access Links.} 
	\label{table_accesslink}
	\centering	 
	\begin{tabular}{|l|  l|  l| }
		\hline
		\textbf{Characteristic feature} &  \textbf{WiFi} & Cellular\\
		\hline
		Bandwidth support & 2.4/5.0 GHz  & Sub-6 GHz (e.g. 2, 3.4 GHz), mmWave (24, 28, 38 GHz)\\ \hline
		Interference Management for cellular operators&Difficult&Easy\\ \hline
		Interference to neighboring cellular links & None & Possible\\		 \hline
		Mobility support &Only deterministic& Deterministic and random\\ \hline
		Deployment cost &Low& Low\\\hline
		Maintenance cost &Low& Low\\\hline
		Connectivity to cellular network &In-direct links&Direct\\\hline
		Provision of in-vehicle data services &Yes&Yes\\ \hline
		Provision of in-vehicle cellular services &No&Yes\\\hline
	\end{tabular}
\end{table*}

In the literature, some researchers proposed to use WiFi-based access links for mobile cells to completely isolate in-vehicle and out-of-vehicle communication frequencies \cite{cook2019mobile}. WiFi-based Internet services are already being offered by many transport providers across the globe \cite{jiang2015bus}. The WiFi transmission uses unlicensed Industrial, Scientific, and Medical (ISM) radio bands that follow CSMA MAC protocols. The in-band interference management in unlicensed bands is challenging and may raise operational and control issues for network operators \cite{cui2017lte}. Additional issues such as authenticating user associations, billing, or providing link security largely remain vague with WiFi \cite{alimi2019enabling}. Furthermore, the WiFi modules inside cellular devices must have access to the device's Universal Subscriber Identity Module (USIM) in order to provide operator-linked services.  This requires hardware and software modifications in the existing device architecture \cite{ganesan2019mobile}. Current WiFi only supports IP-based data services and many cellular operations (such as calling and texting) may not be catered using this technology. For this reason, an on-board WiFi may not boost the actual capacity of cellular networks. Besides, as cellular operators are currently lowering the cost of their data packages, it is anticipated that more users will shift to cellular data as compared to relatively less ubiquitous WiFi technology \cite{yutao2016moving}.  

Due to the above-mentioned reasons, current research works leverage cellular access links within a mobile cell. The in-vehicle transmissions for access links use very low powers because of considerably small  transmitter-to-receiver distances. Hence, the distance dependent path losses significantly decrease. In \cite{jaffry2019efficient}, the authors utilized this fact to demonstrate the sharing of frequency resources between uplink access link and the out-of-vehicle transmissions. The authors showed that using an efficient power control mechanism and maintaining a minimum distance constraint mobile cell links can share resources between multiple links. These links included uplink access link for in-vehicle user, sidehaul link, and uplink access link for out-of-vehicle macro-cell user. The authors also proposed algorithms that allow resource sharing with macro-cell cellular users that are farther away from the mobile cell and closer to the macro-cell base station. The authors reported that ergodic rates for the resource sharing links are very high. For example, both algorithms achieved rates of more than 12 nats/sec/Hz (1 nat  $\approx$ 1.443 bits). Similarly, authors in \cite{jaffry2019interference,jaffry2018shared} also reported the benefits of resource sharing between cellular access link with out-of-vehicle links. 

The mmWave-enabled cellular band can also be a potential candidate for in-vehicle communication in moving networks. For example, the authors in \cite{mastrosimone2015new} presented the feasibility for using 60 GHz access link inside mobile cell with LTE backhaul to form a hybrid mechanism. The high penetration rate of mmWave transmission is a main deterrent to its deployment for cells that provide wider coverage for indoor-to-outdoor transmission services \cite{hattori2015study,niu2015survey,mastrosimonemoving}. However, this characteristic makes mmWave an appealing prospect for smaller sized in-vehicle communication \cite{marzi2015interference}. The mmWave access links also eliminate interference with the sub-6 GHz frequency bands for out-of-vehicle communication. Furthermore, the mmWave communication require highly directed transmission which means that the interference to the unintended directions will be very weak. Highly directional transmission, coupled with poor penetration property of mmWave band, enable frequency reuse even inside the neighboring mobile cells. Note that mmWave band have wide bandwidth available as compared to sub-6 GHz band. However, the cost per sub-channel still remains high for network operators. Hence this frequency reuse will also benefit mobile operators from cost point of view.  

In Table \ref{table_accesslink}, we present a comparison of two competing access link technologies presented in this section.  In the next section, we discuss the existing solutions pertaining to enhanced performance through moving networks, interference management, and handover in moving networks.

\section{Survey of the Current Solutions in Moving Networks}
\label{current_sol}
In this section, we discuss the current solutions in moving networks from the perspective of enhanced performance, interference management, and handover in moving networks. First, we discuss the solutions in the literature that show the performance enhancement of cellular networks through moving networks. Then we discuss a very important issue of the interference management in moving networks, its rationale, and current available solutions in the literature. Finally, we discuss the handover issue in the moving networks and current efforts by the research community to address these challenges. We thoroughly survey the existing solutions. It is important to note that covering every aspect of the moving network is not feasible for one paper, and therefore, we focused on the aspects of the moving networks that play pivotal role in the realization and success of the moving networks deployment. At the end, we derive and discuss valuable lessons from the surveyed solutions.

\subsection{Performance Enhancement with Moving Networks}
\label{sec_perf}
A major challenge for telecom operators is to devise mechanisms to boost the performance of networks. While moving networks can enhance network performance by  facilitating the core network and offloading the macro cell base station, there are several associated challenges. These challenges are mainly linked to the mobile nature of moving network entities (buses or trains etc.). We discuss these challenges and the proposed some solutions in this section. We sub-categorize the most common performance parameters as following: network capacity and coverage, throughput and QoS increment for cellular users (both in-vehicle and out-of-vehicle), and bottleneck at the moving network's backhaul. \linebreak

\subsubsection{\textbf{Network Capacity and Coverage}}
\label{subsec_networkcapacity}

\begin{figure*}[t!]\centering
	\includegraphics[width=5 in, trim={4.0cm 6.0cm 4.5cm 4cm},clip = true]{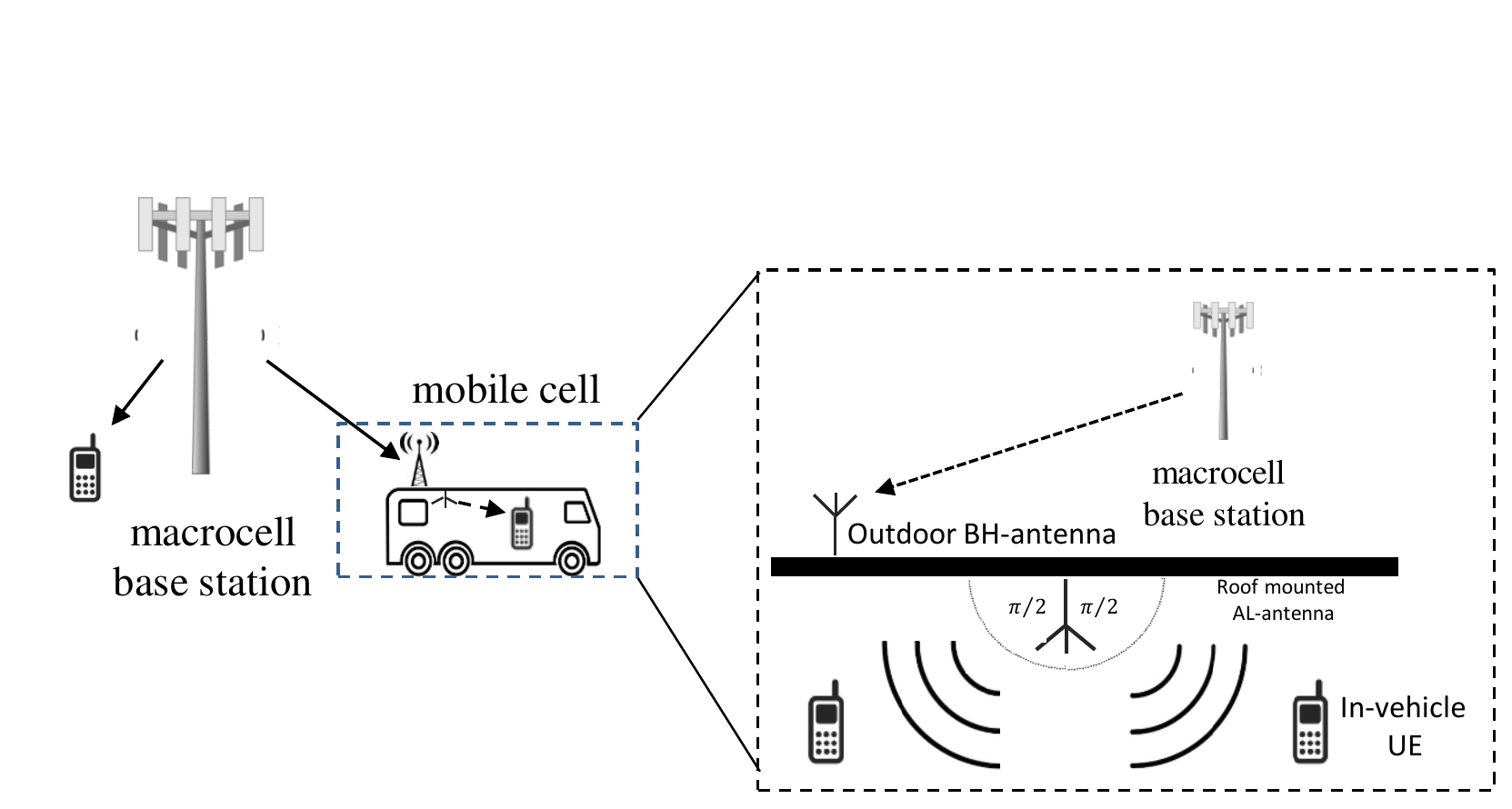}
	\caption{Mobile cell backhaul and access link antenna configuration as shown in  \cite{jaffry2019interference,jaffry2018shared}. The indoor and outdoor (vehicular) antennas are separated by the vehicle's roof (body). The in-vehicle antenna is mounted such that its viewing angle is $\{\pi/2, \pi/2\}$. This orientation ensures that only minor sidelobe signals can interfere with the backhaul antenna, which are reduced in strength by VPL. }	
	\label{fig:jaffrymobilecell}
\end{figure*}

Network capacity is a broader term which means the amount of users a network can accommodate or the available bandwidth in a network. From either definitions' perspective, increasing the network capacity is becoming a key challenge for service providers due to continuous growth in the number of cellular users and traffic demands. Current wireless capacity seems exhausted with overcrowded channels and novel measures are being considered for capacity growth \cite{li20145g}. In 5G networks, operators aim to respond to the increasing traffic demand by dense deployment of small cells and offloading communication to D2D links \cite{jaffry2017distributed} to increase the frequency reuse. Furthermore, as most frequency bands below 6 GHz are already occupied with excessive wireless communication, researchers have been exploring the scantly used high frequency mmWave bands. Although mmWave has very wider contiguous bandwidth, the cost per sub-channel is still high \cite{nikolikj2014profitability, GSMA_report2018}. Hence, service providers are striving to improve spectral efficiency and spectrum reuse to accommodate an ever increasing number of users with limited number of channels. 

To boost the network capacity, resource sharing with in-band relays can be an intuitive and efficient solution. In such a relay system, a single sub-channel is shared by backhaul and access link in the time domain. For example, in \cite{sui2013energy,sui2012potential,sui2012performance}, the authors proposed sub-channel sharing in mobile relays. In the proposed schemes, a sub-channel $f$ was used by the backhaul link in the first LTE sub-frame. The access link reused the same sub-channel in the second sub-frame. The problem with these approaches is that they do not increase the spectral efficiency as the resource sharing was split in time to avoid the high levels of interference between backhaul and access link antennas..

In \cite{jaffry2018shared,jaffry2019interference}, Jaffry et al. demonstrated the sub-channel sharing between in-vehicle and out-of-vehicle links in the same sub-frame. The authors  exploited vehicular penetration losses to enable resource sharing with minimal interference. The authors used a low powered and highly directional AL-antenna for in-vehicle transmission as shown in Figure \ref{fig_mobile_cells} and Figure \ref{fig:jaffrymobilecell}. Due to the directional AL-antenna, the OV-antenna (for backhaul) experienced very low interference from the in-vehicle communication. With a thorough mathematical analysis and simulations, the authors reported high success probability of resource sharing links. Improved  ergodic rates for in-vehicle users and the backhaul links were also reported. It is worth noting that the macro-cell base station does not always transmit backhaul signals to the mobile cells. Hence, in the absence of backhaul signals, \cite{jaffry2019interference} proposed algorithms to extend sub-channel sharing between in-vehicle access link and of out-of-vehicle macro-cell users' access link in \cite{jaffry2019interference}. The authors further used Successive Interference Cancellation (SIC) technique at the backhaul antenna to further reduce the in-band interference from spatially closer AL-antenna. An algorithm, supported by mathematical analysis, was presented to ensure that the resource sharing macro-cell users are selected in such a way that they are closer to the base station, while at the same time farther away from the the mobile cells with which they share the resource. The authors reported a drastic improvement in the success probability of resource sharing links using  SIC and VPL. 

\begin{figure}[t!]\centering
	\includegraphics[width=\linewidth, trim={0.0cm 0.0cm 0cm 0cm},clip = true]{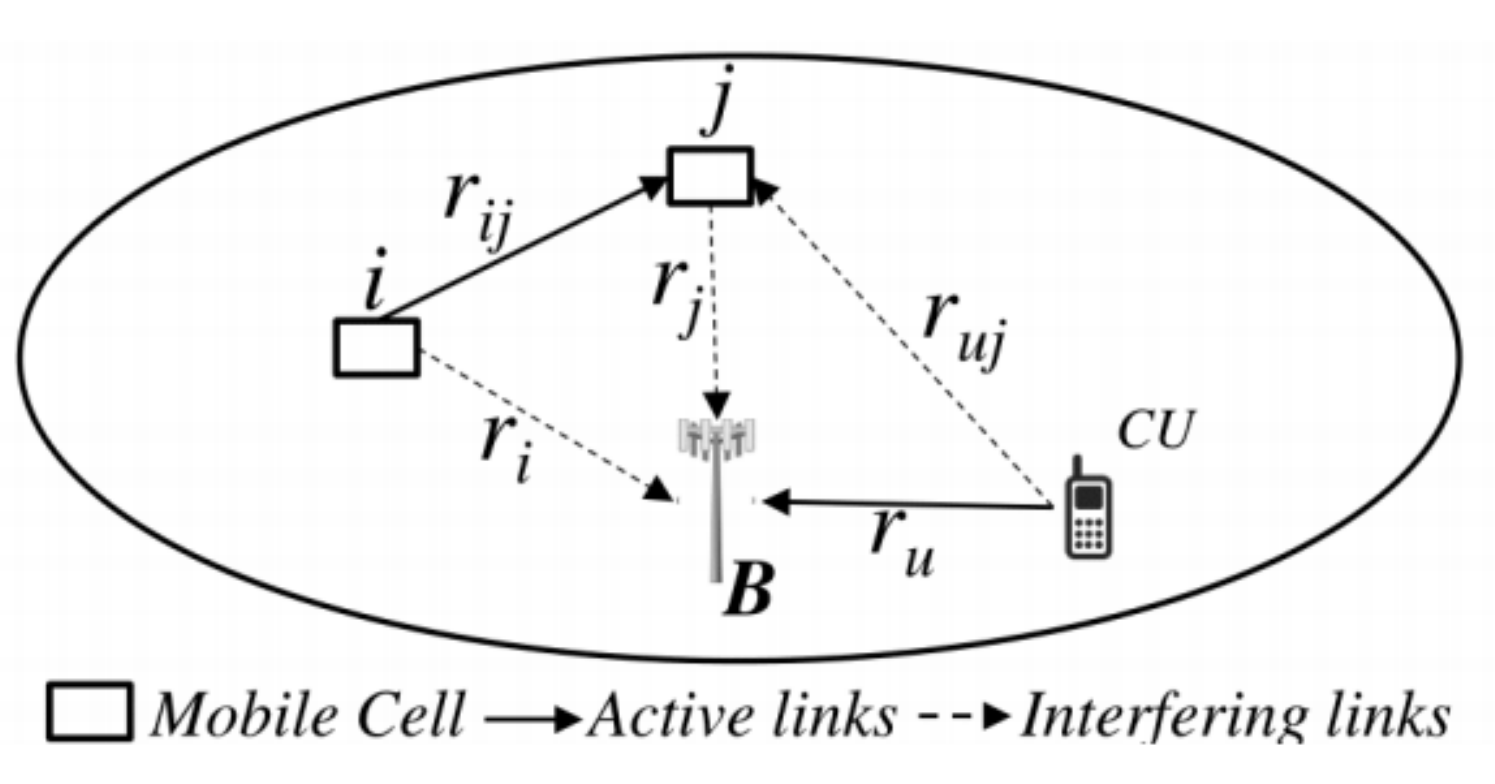}
	\caption{Resource sharing among 'trio' in \cite{jaffry2019efficient}}. $i,j$ are the mobile cells with $i$ sending sidehaul transmission to $j$ on some frequency $w$. Mobile cell $j$ also uses $w$ for uplink access link transmission. Meanwhile, CU shares $w$ with $i,j$ for uplink macro-cell transmission to eNB $B$.
	\label{figch1:jaffryefficient}
\end{figure}

In \cite{jaffry2019efficient}, the authors proposed aggressive resource sharing algorithms in a mobile cell environment. The proposed algorithms focused on enabling resource sharing among following three links  (i) mobile cell uplink access link, (ii) macro-cell uplink access link, and (iii) sidehaul links between the mobile cells as shown in Figure \ref{figch1:jaffryefficient}. The so-called 'trio' of links was selected by the macro-cell base station. The authors observed that the optimal node selection for the trio would require brute force search which is known to be time consuming and impractical in high mobility scenarios involving mobile cells. On the other hand, random allocation of resources yielded very low ergodic rates for all resource sharing links. Hence, the authors proposed two sub-optimal heuristic algorithms to form resource sharing trio with a constraint on maximizing ergodic rates for all links. The authors also reported a strong impact of transmission power on the ergodic rates. However, the selection of transmission power was left as an open question for future research.

In Section \ref{sec_usecases}, we already discussed that mobile cells will provide services to out-of-vehicle users via sidehaul links. In fact, mobile cells can extend the cellular range to the regions with low or no network coverage \cite{shah2018moving} as shown in Figure \ref{fig_MC_out_of_cov}. Particularly in \cite{khan2016moving}, the authors reported cellular coverage extension with mobile cell's cooperative communication via sidehaul link. Similarly, the authors of \cite{MShinPublicSafety} also presented mobile cell aided range extension for cellular networks in the unconnected regions. In \cite{jaffry2018mobile}, authors proposed the idea of using moving networks to provide connectivity to mobile black spot regions, i.e., regions with poor cellular coverage where it is difficult to build an infrastructure due to to difficult terrain or low return on investment for mobile operators. On the other hand, Khan et al. \cite{khan2017outage} showed that the installation of dedicated antenna within trains could increase uplink cellular coverage for on-board users by 50\%. Similarly in \cite{park2019open}, the authors proposed a power control scheme for moving networks to ensure better QoS and reliability for the cell edge users. Along with expanding the coverage and capacity of cellular networks, mobile cells will also increase the throughput and QoS for cellular users. \linebreak

\subsubsection{\textbf{Throughput and Quality of Service}}
\label{subsec_throuhput}

Mobile cells will improve the throughput and QoS for in-vehicle users, along with providing services to the out-of-vehicle users associated with the static layer, for instance in the densely populated urban areas \cite{jaziri2016offloading}. Note that static layer includes fixed macro and small cells. Activities that observe huge public gatherings often exert unusually high traffic demand on the static cellular layer. For example, participants in sports or major public events tend to broadcast live video streaming or regularly update their status on social media platforms \cite{parwez2017big}. Similarly, traffic jams in metropolis often witness a spike in cellular and data traffic due to cramming of large number of users in a relatively smaller area. The excessive resource utilization under such circumstances often causes outages due to collisions. Such collisions occur when more number of cellular users try to access limited amount of available resources. A legacy solution to maintain normal operations in such high density regions is to install a myriad of fixed small cells and use them on-demand. An obvious weakness in such approach is the under utilization of infrastructure when data demands are low. Mobile operators feel reluctant to make investments in infrastructure where the return on investment is low (as in the above-mentioned scenario). The non-static nature of moving networks makes them an ideal and cost effective alternative as compared to fixed small cell infrastructure as discussed in  \cite{marsan2019towards,jaziri2016offloading,nakayama2020small, mohammadnia2019mobile,jiang2015spatio}.

In \cite{jaziri2016offloading}, the authors simulated an environment in which mobile cells provide services to in-vehicle users as well as to the out-of-vehicle users in hotspot regions, to enhance network throughput. The authors reported that the radio conditions of the  hotspots improved when mobile cells get closer to those areas. Hence, macro-cell could offload users to the mobile cells in such cases. It was also reported that due to this offloading, the overall network load decreased and throughput increased drastically. The authors, however, cautioned that a controlled mobility is required to benefit from the moving networks and random mobility (i.e., uncontrolled mobility) in fact reduce the network throughput. 

In \cite{tsai2019throughput}, the authors compared the orthogonal and non-orthogonal resource partitioning in moving networks. The authors studied a network environment with macro-cell, fixed small cell, and mobile cells. Focusing on system throughput and link reliability for mobile cells' resource partition, the authors reported interesting results. The authors reported an increase of more than 62\% in average cell throughput with non-orthogonal resource partitioning compared to orthogonal partitioning with a simultaneous deployment of seven mobile cells. 

In \cite{nakayama2020small}, the authors found that the throughput improvement was roughly 60\% greater when mobile cells were static as compared to when they were moving at a speed of 40 Km/h. Based on this observation, new use-cases of moving networks were explored in other researches \cite{marsan2019towards, mohammadnia2019mobile,nakayama2019adaptive, nakayama2019experimental, nakayama2020small}. 

In \cite{marsan2019towards}, the authors observed that the cellular traffic follows a highly fluctuating but periodic pattern in urban areas. For example, they found that cellular traffic demands in the business districts remain very high during office hours as compared to the night time (which is a logical argument). Similarly, the traffic demands in the residential areas peaked during evenings and the weekends (again a logical argument). This fluctuating patterns were found to be highly correlated with the vehicular mobility patterns in metropolis. The authors used real-world mobility and cellular data from Milan city in Italy to substantiate their claim. In particular, the authors observed mobility patterns of vehicles moving from residential areas to the business districts (and back) in Milan. A surge in vehicle (and subsequently users) flow was found to move from residential areas to business districts. These users travel back to their homes in residential areas in the evening. Consequently, cellular demands regularly shifted between residential areas and the business districts. In conclusion, the mobility of cellular users influence the cellular traffic demands and both quantities were found highly correlated. As mentioned earlier, the legacy approach to satiate the cellular demand is to install more small cell base stations throughout business districts and residential areas. The installation of more base stations increases the throughput by exploiting aggressive frequency reuse. However, installing large amount of fixed base stations also increases the CAPEX and OPEX for the network operators. Furthermore, during the off-peak hours (for example, during the night time or weekends in business districts), these installations are underutilized due to low cellular activities.  

To address the above issues, different researchers proposed to  install base stations on the vehicles that frequently move back and forth between regions of fluctuating cellular demands, such as, between homes and offices \cite{marsan2019towards, nakayama2019adaptive, nakayama2019experimental, nakayama2020small}. This way the moving network travels '\textit{with}' the cellular users. In \cite{marsan2019towards}, the authors reported a drastic increase in the cellular network throughput using the proposed scheme as opposed to installing additional fixed small cells. The authors reported that the static small cells could provide only fixed average network throughput. On the other hand, the throughput increased roughly from 200 Mbps (for static macro and small cell deployments) to a maximum of 700 Mbps with the installation of antennas on vehicles parked in office parking lots. Similarly, in \cite{nakayama2020small}, the authors found a strong correlation between cellular traffic demand and occupancy rate of neighboring parking lots similar to \cite{marsan2019towards}. The authors proposed to install cells on the vehicles that are frequently parked in parking lots (near offices or shopping plazas etc.), to address the surge in data demand during office hours. The authors presented the basic idea in \cite{nakayama2019adaptive} and \cite{nakayama2019experimental} and reported comprehensive analytical, simulation, and experimental results in \cite{nakayama2020small}. In their scheme, the authors named the vehicle-mounted base stations as on-demand Crowd-sourced Radio Units (CRUs). In the simulations, the authors compared three cases. Case 1 only had a macro-cell base station, case 2 had macro-cell with the dense deployment of fixed small cells, while case 3 had macro-cells with CRUs. It was reported that $5^{th}$ percentile throughput for case 2 improved by 39\% as compared to case 1. The throughput for case 3 improved by 58.5\% as compared to case 2 near the cellular traffic hotspots. The authors also noted that replacing static small cells with CRUs also reduced the number of base stations by at least 25\%. Hence moving networks (such as CRUs) can save huge costs for cellular operators. 

\subsubsection{\textbf{Backhaul Bottleneck}}

In any dual-hop network such as mobile cells or relays, the backhaul link may become the main bottleneck \cite{jaber20165g}. Hence, it is imperative to expand the capacity and improve the throughput at the backhaul link. In \cite{sui2014deployment}, the authors proposed to use a multi-antenna system with {\it maximum ratio combining} and {\it interference rejection combining} techniques to improve reception at the backhaul antenna. More recently, mmWave-enabled backhaul was also studied for mobile cells to provide multi-Gbps connectivity \cite{song2016millimeter}. Although mmWave offers a wide contiguous band, it has inherent physical challenges. For example, the mmWave transmission has inferior propagation characteristics and experiences high free space path loss following the Friis' law \cite{heath2016overview, theodore2002wireless}. In addition, as the Doppler shift/spread is proportional to the carrier frequency, the Doppler-induced problems for high speed vehicles get worse for mmWave-enabled backhaul \cite{herbert2014characterizing}. Moreover, the mmWave signals also suffer from additional losses due to poor diffraction, atmospheric absorption, rain attenuation, and high penetration loss \cite{rappaport2013millimeter, andrews2016modeling}.

Researchers have proposed different solutions to address inherent mmWave-band challenges for moving network applications. For example, in \cite{noh2019realizing}, authors proposed to install several mmWave-enabled radio units (RUs) along the roads or the railway tracks. The RUs were responsible for mmWave transmissions to the mobile cell and were connected to the data unit (DU) through radio-over-fiber links. The DU was located in a centralized location, such as, near the train station. These RUs and DUs collectively provided the functionality of a base station. For example, the DU performed higher layer computations and processing, while RU communicated with the mobile cells' over mmWave backhaul links. The commuting users inside the mobile cell accessed the network through an on board relay deployed inside transportation vehicles. The authors used a 25 GHz mmWave frequency band for backhaul with OFDMA transmission following time division duplexing (TDD) for uplink and downlink. The 2x2 dual polarized MIMO antennas were used for the backhaul antennas and the RU. The tests were conducted on real transport in Seoul railway and buses in South Korea. For the railway tests, four RUs were installed between station-1 and station-2\footnote{Station-1 was Jamsil station, station-2 was Seokchon station, and station-3 was Songpa station. All these stations were located in Seoul, South Korea.} to cover the curved path between the stations. One RU was installed near the station-3 which covered the entire straight path. The locations of these five RUs were determined after thorough wave propagation tests. After RU deployment, the downlink data rate of 1.25 Gbps and uplink rate between 550 - 600 Mbps were achieved. As for the bus connectivity, the link performance test achieved a maximum data rate of 5 Gbps and average rate between 2 to 4 Gbps with an aggregated bandwidth of 1 GHz. 

\subsection{Interference Management}
\label{subsec_interference}

\begin{figure}[t]\centering	
	\includegraphics[width=3.5 in, trim={6.0cm 4.5cm 4.5cm 4.0cm},clip = true]{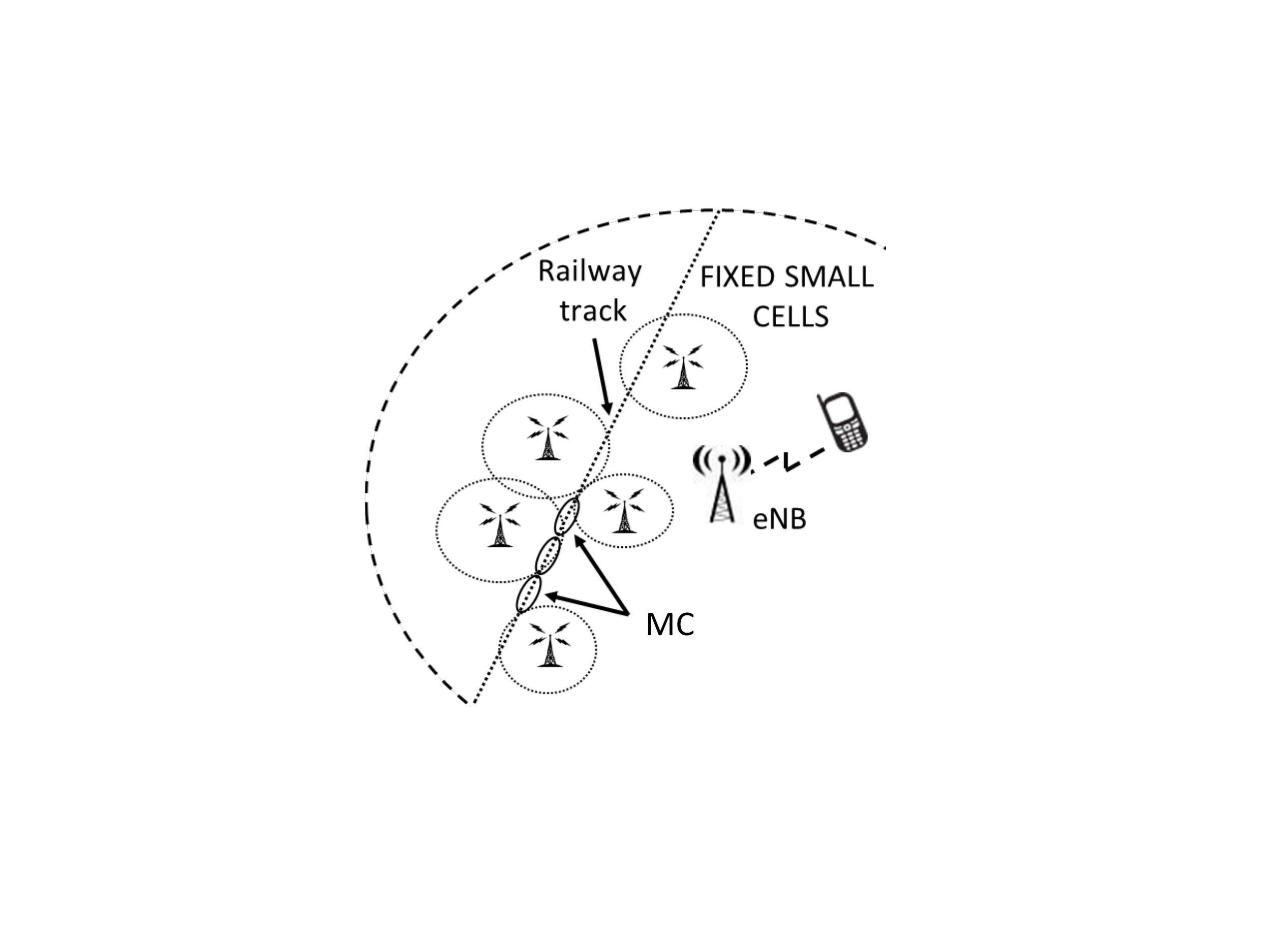}
	\caption{System model as presented by Jangsher et al. for deterministic mobility \cite{jangsher2013resource} and \cite{jangsher2015resource}. }
	\label{S_Jangher_MSC}
\end{figure}

The random and non-stationary layout of moving networks is entirely different from any pre-planned fixed cellular architecture. This randomness causes  serious issues related to interference. Two main types of interference are inter-moving-network interference (i.e., interference among neighboring mobile cells, or between mobile cells and its users), and interference between moving networks and fixed network layer. In this context, the approaches investigated for fixed small cells are not suitable for moving network interference management and separate investigation is needed in this regard \cite{jaffry2016making}. 

\begin{table*}[!ht]
	\renewcommand{\arraystretch}{1.3}
	\caption{Summary for existing solutions dealing with interference management.} 
	\label{table_interference}
	\centering
	 
	\begin{tabular}{|l | p{6cm} | p{6cm} |}
		\hline
		\textbf{Articles} &  \textbf{Mitigation technique(s)} & \textbf{Radio technology for  access and backhaul links}\\ 		\hline
		\cite{jaziri2016offloading}& Analysis on planned v/s unplanned mobility & LTE-A (same band for in and out of vehicle links)\\ \hline
		\cite{hsueh2011novel} & Hybrid optical and wireless system with radio-over-fiber for backhaul & Radio-over-fiber backhaul, any access link. \\ \hline
		\cite{jangsher2017backhaul} & Interference management and resource allocation for backhaul & Different frequencies for backhaul and access links \\ \hline
		\cite{jaffry2018shared} & Resource sharing exploiting VPL & Same \\\hline
		\cite{jaffry2019efficient,jaffry2019interference} &  Exploitation of VPL and using SIC. The authors proposed algorithms for selection of links that share resources with the access link & Same\\ \hline
		\cite{chae2012dynamic} & Interference mitigation by using separate bands for in-vehicle and out-of-vehicle users. & N/A\\ \hline
		\cite{jangsher2014resource,jangsher2015resource,jangsher2013resource} & Resource allocation using mixed integer integer programming and graph coloring& N/A \\ \hline
		\cite{zafar2019resource} & Resource allocation using LSTM ad graph coloring & N/A \\ \hline
		\cite{mastrosimone2015new} & wireless backhaul for mobile cells coupled with mmWave access link to isolate the in-vehicle and out-of-vehicle interference & mmWave for access link, LTE-A for backhaul  \\\hline
		\cite{park2018flocking} & Fairness in resource allocation through power control using Cucker-Smale flocking model& N/A\\ \hline
	\end{tabular}
\end{table*}

In \cite{jaziri2016offloading}, authors used real-world parameters to demonstrate that the unplanned mobility of mobile cells can significantly deteriorated the QoS for both in-vehicle and out-of-vehicle users when both share the same frequency spectrum. However, it was also shown that the planned  mobility for mobile cells could significantly increase the network throughput and offload the macro-cell base station.

To mitigate interference, a simplistic approach was proposed in \cite{chae2012dynamic} in which authors proposed to separate in-vehicle and out-of-vehicle spectrum. However, the authors did not provide extensive analysis or network level simulation to substantiate their claim. A major argument behind moving networks is that they will improve spectral efficiency through frequency reuse which cannot be achieved through a rudimentary approach. However, we will discuss later on in this section that some researchers have argued to use mmWave band in-vehicle and non-mmWave band for out-of-vehicle communication (and vice-versa).  

To eliminate in-band interference in moving networks, a more comprehensive and optimization based resource allocation scheme was proposed by Jangsher et al. in \cite{jangsher2014resource, jangsher2013resource, jangsher2015resource}. The authors combined graph theory and Mixed Integer Non-Linear Programming (MINLP) to solve resource allocation problem. In \cite{jangsher2014resource}, the authors proposed probabilistic resource (i.e., sub-channel and transmit-power) allocation for mobile cells using public transport mobility model. The proposed probabilistic model was based on the presumption that the buses often follow non-deterministic route and timing. For vehicles that have deterministic mobility, such as trains, resource allocation algorithms were presented by the same authors in \cite{jangsher2013resource} and \cite{jangsher2015resource}. 

\begin{figure}[t]\centering 	
	\includegraphics[width=3.0in]{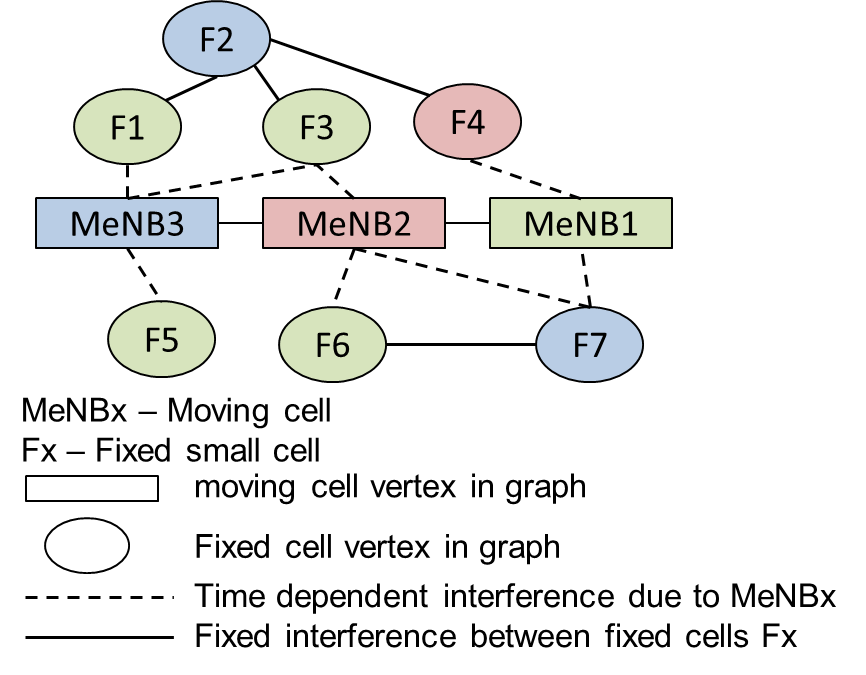}
	\caption{Graph coloring as presented by Jangsher et al. for deterministic mobility \cite{jangsher2013resource} and \cite{jangsher2015resource}. }
	\label{S_Jangher_coloring}
\end{figure}

 In the proposed schemes in \cite{jangsher2014resource, jangsher2013resource, jangsher2015resource}, a centralized controller (e.g., base station) calculated the interference pattern for fixed time slots. Due to their mobility, mobile cells interfere with adjacent cells (both mobile and static) as they travel through their routes. Hence, interference patterns change with respect to time which was calculated by the the central core. The unit of time in these was the time slot. For each time slot, several sub-slots were formed based on the consequent interference patterns, i.e., a single Time Interval Dependent Interference (TIDI) graph per sub-slot. The number of TIDI graphs depends on the speed and direction of vehicles. The core solved the optimization problem for sub-channels (also power in \cite{jangsher2015resource}) allocation for up-link transmission for each TIDI graph in a single time slot. In summary, the centralized core performed the following tasks: (i) formation of graph for a time frame, (ii) solving the joint optimization problem for sub-channels and power allocation, and (iii) resource distribution using graph coloring. Figure \ref{S_Jangher_MSC} shows the system model and Figure \ref{S_Jangher_coloring} shows graph formation. The solid lines in Figure \ref{S_Jangher_coloring} shows the permanent interference between two cells, e.g., between the neighboring fixed cells or the mobile-cells. The dotted lines indicate varying interference between the cells for a given time period. Intuitively, fine tuning for time slots is required to achieve accuracy which exerts additional load on the base station in terms of computational complexity. 
In \cite{jangsher2017backhaul}, the same authors presented the resource allocation for backhaul link using separated spectrum for in-vehicle and out-of vehicle link. 
In \cite{zafar2019resource}, Zafar et al. investigated the resource allocation for moving networks considering multiple inter city bus routes. The authors argued that resource allocation for mobile cells require the exact knowledge of vehicle's location which is trickier to obtain due to mobility. The authors used Long-Short-Term-Memory (LSTM) neural networks to determine the future locations of mobile cells to form an interference pattern. These predictions were made during offline mode. The authors made a Percentage Threshold Interference Graph (PTIG) - a variant of graph presented in \cite{jangsher2014resource}. The resources were allocated using greedy graph coloring scheme to the PTIG graph vertices. Using this scheme, authors reported a 13.50\% decrease in resource block assignment - i.e., an improvement in spectral efficiency - as compared to normal graph used in \cite{jangsher2014resource}.  

The spectral efficiency of any interference management scheme is of prime importance due to scarcity of resources. Authors in \cite{jaffry2019efficient,jaffry2018shared, jaffry2019interference} proposed to exploit VPL and used SIC technique to enable resource sharing in moving networks instead of assigning separate resources for multiple links. In \cite{jaffry2018shared}, the authors focused on sharing mobile cells' downlink access link resource with the backhaul link. In \cite{jaffry2019efficient}, authors enabled resource sharing between access link and out-of-vehicle user in the downlink transmission. Similarly, in \cite{jaffry2019interference}, the authors focused on simultaneous sharing of spectral resources between three links, i.e., mobile cell's up-link access link, sidehaul link, and out-of-vehicle macro cell user. The authors focused on reducing the interference levels on the respective resource sharing link with the exploitation of VPL and SIC. 

A Cucker-Smale flocking model based transmission power control scheme was designed in \cite{park2018flocking} for fair bandwidth allocation in mobile relays. In \cite{liang2019dynamic}, the authors proposed interference-weighted clustering algorithm to extract the interference patterns of mobile cells and proposed the dynamic resource allocation algorithm for heterogeneous networks with mobile cells. On the other hand, cross-tier interference for mobile cells using statistical channel estimation with mobility consideration was presented in \cite{byun2019cross}. In Table \ref{table_interference}, we summarize the existing works that discussed interference management issues in moving networks.

Another major issue concerning high speed vehicles in moving networks is handover. Next, We discuss the solution proposed for handover problems.

\subsection{Handover}

In cellular networks, handover is required to support user mobility when they travel from the coverage of one base station to another \cite{lee2018performance}. Legacy handover mechanisms are based on comparing  received signal strength, i.e., a handover is triggered when the received signal strength of the target base station exceeds that of the serving base station by a certain threshold offset \cite{park2018handover}. These kinds of handover triggering schemes may be suitable for conventional cellular networks. However, in an environment with an abundance of fast moving transport vehicles, commuters-induced group handovers may cause excessive overhead on the network \cite{merwaday2016handover}. For example, a bus traveling with a modest speed of 50 km/hr can cover a distance of nearly 1 km per minute. During this 1 Km, a vehicle may need to handover multiple times depending on its traveling path. If the bus is loaded with 50 passengers, all  simultaneously using the mobile phones, it means that the network has to deal with 50 simultaneous handovers. The situation gets worse for fast moving trains which may take speeds of up to 500 km/hr in which case they may travel 1 km in merely 7.2 seconds. A regular train may contain more than 100 passengers per railway van which may exert excessive load on the network due to large number of group handovers. 

The  mobile cells or relays inside public transport will reduce the group handovers to a single handover \cite{sui2012performance, hsueh2011novel}. This will be  achieved by detaching the commuting users from the core network and instead serving them with an intermediary in-vehicle access link layer. In \cite{nakayama2018optically, nakayama2017optically}, the authors proposed optical fiber backhaul link for fast moving trains and subways which reduced the handover count downs to zero. This was due to the fact that the commuters were linked to the in-vehicle antennas. The link to the backhaul network, which is the main reason behind handovers in mobile cells, was provided by the optical fiber cable which stretched from one end to the other above the railway line. Other techniques such as separation of control and user planes for in-vehicle users were used to reduce the number of handovers in \cite{yasuda2015study}. 

In \cite{chae2013novel}, the authors utilized the LTE-A Coordinated Multi-Point transmission (CoMP) feature to design a handover mechanism for a fast moving trains equipped with mobile femto cells. CoMP utilizes transmission from the adjacent cells to enhance the performance of the cellular users, especially those located at the cell edges, by coordinating with the neighboring base stations. In \cite{chae2013novel} authors proposed to install each train compartment with individual indoor access points. The handover mechanism  triggered as the train moved from the coverage of its serving cell to target cell. The authors proposed that as the front compartment undergoes handover-shift, the end-compartment of the trains remain connected to the serving base station. The train was installed with a central controller (CC) that connected all the individual femto cells. The first compartment remain connected to the serving base station using the CC over wired interface eliminating the chances of transmission loss in case of handover failures.
 
Handover issues are much severe for mmWave backhaul transmission owing to smaller coverage at high frequency bands. The mmWave backhaul would require multiple radio units installed at short distances along with the railway track or bus routes. This will improve the line-of-sight reception for the mobile cells but does not alleviate the handover issue, in fact it may make it worse. The researchers in \cite{noh2019realizing} proposed a modification in 5G NR frame structure to facilitate handovers in mmWave-enabled backhaul. The main modification was done in the synchronization signal block to enable rapid synchronization with the target mmWave-enable radio unit. With this modification, the handover interruption time - i.e. the shortest time duration during which a mobile device cannot exchange user plane packets due to mobility procedure - was reported to be only 4.75 ms for bus trials. 

\subsection{Lessons Learned}

In this section, we surveyed and discussed the existing solutions in the literature for different aspects of moving networks. These solutions will collectively improve the cellular network performance in the context of moving networks. In the light of these solutions, we derive the following lessons.  

Moving networks play a decisive role in decoupling the in-vehicle users from the core cellular networks. This will benefit the future networks in two very important ways. First, a large number of cellular users will be offloaded from the core networks which will relieve base stations to accommodate more users outside the public transport. In essence, Internet activities and video streaming account for most of the online activities in modern mobile phone era. Hence, the installation of in-vehicle cache inside mobile cells will further reduce the consumption of backhaul bandwidth between the core network and the vehicles. To this end, network operators must design efficient cache updating mechanisms to leverage this benefit from the moving networks. The integration of mobile cells in the cellular architecture will ensure that the core network communicates with fewer mobile cells as opposed to several in-vehicle public transport users. Hence, overall \textit{network to end-device} links will reduce which will consequently reduce the number of handovers in the network. However, handovers for mmWave-enabled backhaul links still require further investigation. Due to multiple wireless links introduced by moving networks (backhaul, sidehaul, access links) interference is also a serious challenge. Researchers have proposed solutions to mitigate interference through resource management techniques.  In addition to these lessons, in the following, we discuss the derived lessons in more detail, from different perspectives.

\subsubsection{\textbf{Dedicated vs Shared Spectrum}} 
\label{subsec_dedic}

The issue of using dedicated or shared spectrum is a direct consequence of our discussion on interference management. The researchers in \cite{chae2012dynamic, kwon2015radio, kerdoncuff2018mobile, tsai2019throughput} used separate bands for in-vehicle and out-of-vehicle communication and reported very high throughput and QoS due to minimal in-band interference. For example, authors in \cite{mastrosimone2015new} used mmWave-enabled in-vehicle access link and LTE-A  backhaul to report high throughput as there was no interference between backhaul and access links.

Cellular spectrum is an expensive commodity and using dedicated spectrum for in-vehicle and out-of-vehicle links is never a desirable option for cellular operators due to poor spectral efficiency. The alternative, i.e., sharing spectrum, has its own set of challenges. Authors in  \cite{jaziri2016offloading} reported that shared spectrum between mobile cell's in-vehicle users and fixed layers' (i.e., macro, small cell) users may cause severe QoS degradation due to high interference, subject to unplanned mobility. However, the authors in \cite{jaffry2019efficient,jaffry2019interference,jaffry2018effective,jaffry2018shared} reported that high link success probabilities and ergodic rates can be achieved using same frequency spectrum for in-vehicle and out-of-vehicle communication, provided that VPL and SIC are utilized. However, it is highly anticipated that network operators may opt for dedicated spectrum for in-vehicle and out-of-vehicle communication due to availability of large contiguous bandwidth and low penetration properties of mmWave spectrum to avoid interference in spatially close AL-antenna and OV-antenna.

\subsubsection{\textbf{Lack of analytical analysis}} 
Most of the existing researches related to moving networks (such as \cite{jaziri2016offloading,shah2018moving, nakayama2018optically, nakayama2017optically,sui2012performance,yasuda2015study, chae2013novel, noh2019realizing,jangsher2014resource,jangsher2015resource,jangsher2017backhaul,jangsher2013resource}) rely on network simulations and experimental setups. The lack of analytical analysis suggests the fact that environment of moving networks is extremely complex due to various factors such as device mobility, dynamic data traffic, channel variations, complex nature of blockages involved in the transceiver paths, and dynamic routes for buses and trains. To date, many resource management problems for moving network environment have been solved as optimization problems with the optimal resource allocation as the objective functions with some constraints such as spectral efficiency \cite{jangsher2014resource,jangsher2013resource}. In most cases, the conventional optimization problems are not solvable due to greater complexity and needed constraint relaxation. In cases where non-convex problems could not be solved optimally, sub-optimal solutions have been derived (such as in \cite{jaffry2019efficient,jangsher2015resource}). In this context, new methods are required to reduce the complexity of problems associated with moving networks.

On the other hand, only few researchers such as \cite{jaffry2019efficient, khan2017outage,khan2016moving, tang2020meta}, used analytical tools such as probabilistic analysis and stochastic geometry and examined the Signal-to-Interference-plus-Noise Ratio (SINR) based coverage or outage probabilities. However, these approaches assume several limitations in their analysis for the sake of analytical tractability and hence deviate from the real-world scenarios. Therefore, the development of proper analytical tools will also support building the strong mathematical foundation for moving network environment. 

\subsubsection{\textbf{Lack of intelligent solutions}}
Another important observation is the lack of Machine Learning (ML)-based or Artificial Intelligence (AI) powered algorithms for solving challenges associated with moving networks. In the current literature, researchers either used conventional network level simulations (\cite{shah2018moving,jaziri2016offloading,nakayama2018optically}), solved issues using optimization problems (\cite{jangsher2014resource,jangsher2013resource,jaffry2019efficient}), or used plain analytical models (\cite{tang2020meta,jaffry2019interference}). To the best of our knowledge, only \cite{zafar2019resource} used LSTM-based neural network to solve interference issues in mobile cell environment. The lack of ML/AI techniques for solving moving networks' challenges outlines the fact that public data related to in-vehicular cellular communication is not readily available which is needed for training any ML/AI model. Anticipating that moving networks will be integrated into cellular networks in an era beyond 5G (perhaps the sixth generation or 6G), the ML/AI based solutions are still an open research arena for future researchers.  

Despite all the value additions of moving networks, it is still in its nascent phase and hence there are some open issues that need more attention before its final integration with future cellular architecture \cite{dresslerdynamic}. In the next section, we discuss these open issues. We also discuss some of the future directions related to moving networks in the next section.

\section{Open Issues and Future Directions}
\label{sec_challenges}
 Moving networks play a pivotal role in the realization of many futuristic services for the consumers. With the inception of ultra-fast and relatively cheaper Internet, moving networks have more room to nurture and help in many enabling technologies such as Internet of Things (IoT) and Intelligent Transportation System (ITS). Despite the current research advances in moving networks, there are still outstanding open issues that are essential to address. We also identify some future directions that still need further investigation.  

\begin{table}	[!t]
	\renewcommand{\arraystretch}{1.3}
	\caption{High Speed Railway around the World as of 2018 \cite{factsheet2019_HSR}} 
	\label{table_railwaydata}
	\centering	 
	\begin{tabular}{|l  |p{3cm}|  p{3cm}| }
		\hline
		\textbf{Country} &  \textbf{Length of operating lines (Km)} & \textbf{Max. operating Speed (Km/hr)} \\
		\hline
		China & 26,869 & 350\\ \hline
		France & 3220 & 320 \\ \hline
		Spain & 3100 & 310 \\ \hline
		Japan & 3041 & 320\\ \hline
		Germany & 3038 & 300\\ \hline
		Sweden & 1706& 205 \\ \hline
		UK & 1377 & 300 \\ \hline
		South Korea & 1104 & 305 \\
		\hline		
	\end{tabular}
\end{table}

\subsection{Challenges in High Speed Railways}
\label{subsec_HSR}

Many countries like China, France, Spain, Germany, South Korea, Japan, and Sweden have developed a widespread High Speed Railway (HSR) network to connect major cities as shown in table \ref{table_railwaydata} \cite{factsheet2019_HSR}. Each year, around 1.5 billion passengers travel through HSR in China alone \cite{chen2018development}. As large number of users travel in a single railway van for extended periods of time, it is not hard to imagine that they would be accessing Internet services, particularly multimedia services. 

\begin{table*}	[!t]
	\renewcommand{\arraystretch}{1.3}
	\caption{Summary of research works exclusively discussing railways, subways, and HSR.} 
	\label{table_trains}
	\centering
	 
	\begin{tabular}{|p{2cm} | p{8cm} | p{3cm} |}		\hline
		\textbf{Article} &  \textbf{Focus} & \textbf{Backhaul RAN}\\ \hline
		\cite{hsueh2011novel} & Hybrid optical and wireless system with radio-over-fiber for backhaul. & radio-over-fiber \\ \hline
		\cite{nakayama2018optically, nakayama2017optically} & optical-fiber links for backhauled in trains. & Optical fiber links\\ \hline 
		\cite{noh2019realizing} & Design of multi-Gbps mmWave backhaul for Trains and buses in Seoul, South Korea. Emphasis on increasing throughput and handover improvement for mmWave transmissions. & mmWave \\\hline 
		\cite{jangsher2013resource} & Interference management and resource allocation in deterministic moving networks such as railways and subways. & LTE-A \\\hline 
		\cite{iturralde2018performance} & Performance comparison when commuters inside trains are served with a on-board relay versus when they receive direct transmission. &  LTE-A \\ \hline
		\cite{feng2017vehicle} & Network throughput improvement through moving network aided communication with out-of-vehicle users. & LTE-A \\ \hline
		\cite{chae2013novel} & Proposed Coordinated Multi Point Transmission for novel handover mechanisms for fast moving trains. & LTE/LTE-A.\\ \hline
		\cite{mastrosimone2015new} & wireless backhaul for mobile cells coupled with mmWave access link to isolate the in-vehicle and out-of-vehicle interference. & LTE-A\\ \hline 
		\cite{song2016millimeter} &multi-Gbps communication design for HSR with special focus on redesigning 5G NR frame structure.& mmWave\\ \hline
		\cite{chen2018development} &Survey on mobile communication system for HSR with focus on emergency messaging. & LTE-A\\ \hline
		\cite{van2010providing} & cooperative relaying mechanism to support in-vehicle broadband. Smooth handover mechanism was proposed. & LTE-A \\ \hline
		\cite{khan2017outage,khan2016moving} & Extension of coverage for in-vehicle users, along with out-of-vehicle sidehaul communication via cooperative relaying. & LTE-A\\ \hline
	\end{tabular}
\end{table*} 

\begin{table*}	[!t]
	\renewcommand{\arraystretch}{1.3}
	\caption{Summary of research works with respect to their subjects.} 
	\label{table_papers}
	\centering
	 
	\begin{tabular}{|l | p{8.5cm} | p{2cm}| p{1.8cm} |}
		\hline
		\textbf{Article} &  \textbf{General focus of paper} & \textbf{Transport type} & \textbf{Transmission technology}\\
		\hline
        \cite{chen2018development} &Survey on mobile communication system for HSR with focus on emergency messaging. & Railway&LTE-A.\\ \hline
		\cite{kim2018comprehensive} & Survey on mmWave enabled communication in railways. & railway & mmWave.\\ \hline 
		\cite{jaffry2016making} & A short survey on interference management in moving networks. & Any & LTE-A. \\\hline
        \cite{wang2015channel} & Survey on channel measurement techniques in HSR. &  HSR & N/A.\\ \hline
		\cite{moreno2015survey} & Radio communication for high speed rails with focus on safety and control message transmission. &  HSR & N/A. \\ \hline
		\cite{fokum2010survey} & Deployment of backhaul and fronthaul network for broadband in railway. & Railways & LTE/LTE-A.\\ \hline
		\cite{tanghe2008evaluation} & Experimental evaluation of vehicular penetration loss in sub-6 GHz band. & Any & N/A. \\ \hline
		\cite{tang2017coverage} & Mobile relay's use of coordinated multi-point (CoMP) joint transmission to serve out-of-vehicle macro-cell users. & Any & LTE/LTE-A. \\ \hline
		\cite{favraud2016toward} & Use of mobile cell for public safety. & Buses & LTE-A. \\ \hline
		\cite{jaffry2018mobile} & General discussion on several use cases of mobile cells & Any & N/A.\\\hline
		\cite{mollah2016novel} & Use of 802.11p for mobile cell wireless backhaul link. & Vans and buses & 802.11p. \\ \hline
		\cite{rodriguez2015lte} & Measurement of LTE downlink performance for trains. & Trains & LTE-A backhaul.\\ \hline
		\cite{jaziri2016offloading} & Evaluation of moving network's mobility effect throughput and QoS of users and network. & Buses & LTE-A.\\ \hline
		\cite{nakayama2018optically, nakayama2017optically} & Optically backhauled links for trains & Trains. & Fiber-optic backhaul links. \\\hline
		\cite{marsan2019towards, mohammadnia2019mobile,nakayama2019adaptive, nakayama2019experimental, nakayama2020small} & Proposed installing base station on vehicles that travel to and from residential areas to business districts or downtown areas to meet fluctuation data demands. & Vans and cars & LTE-A. \\ \hline 
		\cite{shah2018moving} & Dual role of a mobile cell, (i) as a relay to connect out-of-vehicle users,  (ii) and mobile cache for in-vehicle users.  & Buses & LTE-A. \\ \hline
        \cite{jiang2015spatio} & Data analysis of web service usage in public transport (buses) in Sweden. Results include data analysis and simulation with focus on bandwidth consumption with and without in-vehicle caches. & buses & LTE-A \\\hline
        \cite{kwon2015radio} & Content distribution for Cache within mobile cell (backhaul/sidehaul). separate in-vehicle, out-of-vehicle bands. & Buses & LTE-A. \\ \hline  
		\cite{zafar2019resource, iftikhar2019resource,jangsher2014resource,jangsher2013resource,jangsher2015resource}& Interference management and Resource allocation in moving networks. & Buses/trains & LTE-A. \\\hline 
		\cite{jaffry2018shared,jaffry2019efficient,jaffry2018effective,jaffry2019interference} & Resource sharing for mobile cell access link with out-of-vehicle links (macro-cell user, backhaul, sidehaul). & Any & sub-6 GHz. \\\hline	
		\cite{jangsher2017backhaul} & Resource allocation for backhaul links using Integer programming optimization.  & Buses/trains & LTE-A. \\ \hline
		\cite{sui2014deployment} &Use of Multiple antenna for backhaul with maximum ratio combining and interference rejection combining technique. & Buses & LTE/LTE-A\\ \hline
		\cite{sui2013moving,sui2012potential,sui2012performance,sui2013energy} & Several performance aspects of mobile relays including transmission power, energy efficiency, placement of antenna etc. & Any &  LTE-A. \\ \hline
		\cite{tang2020meta} & Meta distribution of signal-to-interference ration for moving networks. & Buses & LTE-A. \\\hline
		\cite{chae2013novel} & Design Coordinated MultiPoint Transmission for novel handover mechanisms for fast moving trains. & Trains & LTE/LTE-A. \\ \hline
		\cite{mastrosimone2015new,mastrosimonemoving} & wireless backhaul for mobile cells coupled with mmWave access link to isolate the in-vehicle and out-of-vehicle interference. & Buses/trains & LTE-A.\\ \hline
		\cite{MShinPublicSafety} & Use of sidehaul link at 700 MHz band to provide cellular coverage to unconnected regions, especially for public safety scenarios. & Buses & LTE-A  \\ \hline
		\cite{yasuda2015study}& Installation of fixed small cell base stations near railway line with massive-MIMO backhaul antenna. The researchers observed the improvement in throughput and reduction in number of handovers control signaling through C/U plane separation. & Trains & LTE-A.\\ \hline
        \cite{tsai2019throughput} & Comparison of orthogonal and non-orthogonal resource partitioning in moving networks. & Buses/Trains & Any.\\ \hline
		\cite{noh2019realizing} & Design of multi-Gbps backhaul for moving networks. Results include simulation, real-world experimental. & Buses/Trains& mmWave. \\\hline
		\cite{song2016millimeter} & Multi-Gbps communication design for HSR with special focus on redesigning 5G NR frame structure. & HSR & mmWave for backhaul, any indoor.\\\hline
		\cite{chae2012dynamic} & Interference management by using separate resource allocation for in-vehicle and out-of-vehicle users. & Buses & Separate bands for in and out of vehicle users.\\ \hline		
		\cite{park2018flocking} & Power allocation for mobile relays using Cucker-Smale flocking model. & Any & Any.\\ \hline
		\cite{byun2019cross} & Cross tier interference for mobile cells with channel statistics estimation.  & Buses/trains & N/A.  \\\hline 
		\cite{wang2018moving} & Securing mobile cell's physical layer communication from eavesdroppers. & Any & N/A. \\ \hline  
	\end{tabular}
	\pagebreak
\end{table*}

HSR trains travel at a speed of 250 - 500 Km/hr to shorten the commuting time for passengers. At this speed, providing low latency and ubiquitous wireless communication is challenging. Due to very high speed, the Doppler effect shifts the operating frequency by a large margin as Doppler-shift is proportional to the speed of the receiver. For example, in Eq. \ref{eq_doppler_formular}, it can be noticed that the velocity ($v$) of vehicle proportionally affects the receiving frequency of radio wave ($f_{obs}$) if the original frequency is $f_{orig}$. Consequently in-vehicle user equipment could not tune with the observed frequency ($f_{obs}$) and may experience call drops. To make matters worse, the metallic body and insulated glass windows of HSR acts as penetrating walls for radio waves which further degrade the signal quality. Using mobile cells inside HSR can circumvent VPL, minimize Doppler-effect, and reduce the number of handovers. 

\begin{equation}
\label{eq_doppler_formular}
    f_{obs} = v f_{orig}\cos{\theta}
\end{equation}

For railway communication, LTE for railway (LTE-R) is a popular wireless communication mode. LTE-R was first deployed in South Korea which can provide speeds of up to 100 Mbps with 20 MHz bandwidth \cite{chen2018development}. However, considering future data demands, which will be mostly driven by multimedia applications, this data rate capacity is very limited. For example, a single high definition video broadband  connection inside a railway van would require data rates of up to 3 Gbps  (allowing 1920 x 1080 \@ 60 Hz \@ 24 bits) on a 40 MHz channel for a single user \cite{ai2015future}. With more than 100 users in a single railway passenger coach, assuming all accessing multimedia services, the numbers add up to an enormous figure from within a single train. 

Hence railway service providers and researchers alike are advocating to provide multi-gigabit transmission for HSR using mmWave band backhaul \cite{noh2019realizing}. However, due to inherent channel hostility towards mmWave band, there are several challenges associated to it. In \cite{song2016millimeter}, authors proposed modifications in the OFDM frame structure for mmWave communication to enable multi gigabit transmission and to alleviate the problem of high Doppler shift. These changes include modifications in the sub carrier spacing, changes in the symbol and sub-frame length, and adding feedback mechanisms.

Another challenge associated with HSR is oftentimes the difficult terrain. Researchers have often struggled with modeling radio channels in complex terrain involving mountains, tunnels, viaducts, rivers, meadows, and urban or rural population \cite{wang2018survey,wang2015channel}. In such a heterogeneous long route for railways, it is a challenge to uniformly provide multi-gigabit transmission at the backhaul. Millimeter wave \cite{noh2019realizing} or fiber optic backhaul \cite{nakayama2018optically} might well suit for underground subways or urban railway tracks, but scaling these models to the very long routes, as mentioned above, is practically very costly. Even with the sub-6 GHz backhaul links, both slow and fast fading effects degrade the link quality at such a high speed, even with highly advanced backhaul antenna \cite{liu2012position}. Therefore, further investigation is needed to develop an understanding of channel models for HSR. Some interesting studies are presented in \cite{zhou2019geometry,zhang2018measurement,zhou2018dynamic} and the references therein.

Furthermore, handover issues for HSR with mmWave backhaul link is highly unexplored topic and need much investigation. In \cite{noh2019realizing}, authors deployed a mmWave-enabled backhaul link for railway in Seoul city, South Korea. Authors  proposed some modifications in 5G NR frame structure to alleviate the handover problem (among other issues) for mmWave backhaul in HSR. Meanwhile, 3GPP has started the standardization efforts for mmWave communication for high mobility use cases as part of 5G NR specification \cite{ahmadi20195g}. It is worth mentioning that in the existing literature, researchers have also examined the essential communication for emergency and control purposes in regular trains, subways, and HSR. This topic is out of scope of the current survey. To study research on HSR related control channel communication, we refer readers to \cite{ai2014challenges}. To study the standardization efforts for 5G NR HSR communication including network architecture, initial access, mobility management, and physical layer design, readers may refer to \cite{hasegawa2018high}. A list of papers that discussed communication in railways is presented in Table \ref{table_trains}.

\subsection{Security}
\label{subsec_security}

Like any other networking technology, security and privacy concerns are of prime importance in moving networks as well. Generally, the research pertinent to security in conventional static layer and Vehicular Ad hoc NETworks (VANETs) can be adopted, with slight variations and tweaks, to safeguard moving networks' communication as well. In particular, VANETs are most closely related to moving networks and hence researchers can borrow some aspects of VANET's solutions in security and privacy to secure moving networks. Readers can refer to some of these solutions in \cite{lu2018survey,alnasser2019cyber,hasrouny2017vanet}, and the references therein. However, it is important to note that most of the research related to VANET security focus on bandwidth-limited ITS which  may not be directly applicable to moving networks and hence further research is still needed in this domain.
To date, some existing literature tries to address the security challenges in moving networks. For instance, in \cite{wang2018moving}, the authors proposed a mechanism to secure physical layer communication for in-vehicle users that use cellular access link. The authors exploited differences in channel state conditions to distinguish between a legitimate in-vehicle user and an eavesdropper (out-side vehicle). The eavesdroppers were denied access to access link transmission by exploiting VPL which severely reduced the signal quality outside the mobile cell. 
For HSR control signals, the security protocols are defined for Global System for Mobile Communications for railways (GSM-R) and LTE-R.  These standards merely deal with the railway's control channels and do not provide security to user communication within HSR. The discussion on GSM-R and LTE-R does no lie within the scope of this survey. Nevertheless, we refer interested readers to \cite{ai2014challenges, zhenhai2010study} etc. for further reading. 

 Despite the current efforts, the security challenges in moving networks will be a bit different and enhanced version of the existing security and privacy challenges in the VANETs. For instance, mobility is the pinnacle of moving networks; however, fast speed introduces extra challenges of ultra-fast authentication, access control, and auditing mechanisms for the service providers. Furthermore, the decoupling of network service providers and other enabling technologies such as IoT will further introduce the need for effective security measures with context in mind. In this regard, context-aware security should be investigated for moving networks. Furthermore, the IP-based backbone of the current cellular technologies also expose them to attacks that are hard to contain. Recently, Machine Learning (ML) and Deep Learning (DL) have been widely used to address security issues in many networks including IoT \cite{Hussain2019}. Further investigation is needed to employ ML and DL for addressing different security problems in moving networks. It is also important to note that the recent enforcement of General Data Protection Regulation (GDPR) will also cause a strain for the service providers to protect privacy of the users and therefore, further research is needed to make the services GDPR-compliant. One possible solution could be introducing transparent data collection and processing algorithms at the core of the service providers; however, there must be a trade-off among the security, privacy, and the investment of the service providers. Therefore, deeper insights are needed in this direction. In a nutshell, security of the moving networks is extremely important and will directly affect the applicability of the moving networks.

\subsection{Energy Efficiency in Mobile Cells}
\label{subsec_power}
In the near future, hybrid or electric powered vehicles are expected to outnumber the fossil fuelled vehicles, at least in the developed economies. These vehicles are partially (or completely) driven by battery power and consume little to no fossil fuels \cite{masuda2019electric}. These vehicles will be the transportation mode in the futuristic green and smart cities. Since future networks (5G and 6G) also envisions a green communication, it is imperative that entities such as moving networks also adhere to the power constraints while not compromising on the QoS. 

The goal of green communication in moving networks should be to use minimal battery power so that primary automotive operations are not hampered. This warrants for designing efficient power control algorithms to provide stronger backhaul, sidehaul, and access link transmissions without draining large amount energy from the vehicle's battery. In this context, research in the directions of energy harvesting \cite{ma2019sensing,imran2019energy} or back scattering for mobile relays \cite{liu2019next,lu2019ambient} are worth exploring. 

\subsection{Practical Considerations} 
In addition to the above-mentioned technical research challenges, a few practical concerns also warrant considerations. For example, commuters inside buses or trains may use services of different cellular operators. The installation of separate in-vehicle equipment (e.g., antennae or cache components etc.) by each operator will require significant deployment and maintenance costs by all the stakeholders (i.e., mobile operators, and transport service providers, etc.). Therefore, effective mechanisms for sharing physical resources such as spectrum, antennae, and caching system, are required to benefit all the stakeholders \cite{afraz2019distributed,zhang2019toward}. This will inevitably encourage the use of Infrastructure as a Service (IaaS) paradigm.

On the other hand, the recent apprehensions that massive amount of 5G base stations will surround our environment, have stirred debates related to constant electromagnetic exposures and its impact on human well-being \cite{wu2015human}. The idea that AL-antennae within mobile-cells will continuously transmit radio signals to spatially closer in-vehicle users may raise more concerns. Therefore, the domain experts should conduct a comprehensive study to regulates the acceptable transmit power-levels which are not only safe for human well-being but also provide high QoS to cellular users.
 
\subsection{Mobile Nodes in the Air} 
\label{subsec_UAV}
An interesting and relatively new topic in moving networks is in-the-air Unmanned Aerial Vehicle (UAV) relays that can be deployed with UAVs \cite{zeng2016wireless}. Though UAV relays do not strictly follow our definition of mobile cells, i.e., they do not have in-vehicle users; however, they can extend services of conventional base station in regions with low or no connectivity. These UAV-relays may also provide backhaul links to the on-ground mobile cells \cite{zeng2016throughput}. The UAV-relays offer a new degree of freedom due to more extensive range of transmission as compared to ground nodes. Authors in \cite{zhao2018caching} explored delivering cached contents to ground users using UAV-relays. An interesting direction of research can be \textit{mobile cell to UAV communication} for backhaul transmissions when the link to macro-cell base station is weak or non-existent. Another important direction concerning UAV communication can be modelling of three-dimensional (3D) channels for UAV assisted communications \cite{jin2017three, jiang2018three}.
 
\section{Conclusion}
\label{sec_conclude}

In this paper, we conducted a a comprehensive survey on the moving networks which will be an integral part of future cellular networks. In moving networks, mobile cell are installed inside public transport vehicles, such as, buses, trains, and subways etc. The commuters inside these transport vehicles often experience low QoS and throughput due to poor signal quality because of vehicular penetration losses. Doppler shifting also plays a major role in deteriorating QoS in high speed vehicles. Furthermore, as public transport vehicles carry large number of cellular users (e.g. more than 100 users in single carrier of train), they have an inherent problem of large group handovers. Consequently these issues contribute in degrading users QoS and network's performance. The in-vehicle antenna inside mobile cells will circumvent VPL and improve the signal quality inside the transport vehicle. The commuters will be decoupled from the core network and an out-of-vehicle backhaul antenna will communicate with the core network. This way large number of group handovers will ideally diminish to a single hand-off per mobile cell. The advanced backhaul antenna supported by mobile cell can also compensate for the Doppler shifting effects. 

We investigated and discussed in this survey, the current literature where researchers proposed interesting use cases and applications of moving networks such as mobile caching, coverage extension to regions with low connectivity, and adaptive radio networking etc. Furthermore, we also discussed moving network architecture, along with major challenges associated with mobile cells that need to be solved before they can integrate with the future cellular network. Some exciting new technological advancements and future directions, such as mmWave-enabled multi-gigabit backhaul transmission for HSR are also discussed in this survey. We anticipate that our survey will not only serve the researchers working in the domain of future cellular networks, but also to the network service providers and public transport operators interested in providing value added services to the commuters.

\bibliographystyle{IEEEtran}
\bibliography{literature_references}

\end{document}